\documentclass[superscriptaddress]{revtex4}
\usepackage{epsfig}
\def\ii{\'{\i}}
\def\beq{\begin{equation}}
\def\eeq{\end{equation}}
\def\beqa{\begin{eqnarray}}
\def\eeqa{\end{eqnarray}}
\def\ban{\begin{eqnarray*}}
\def\ean{\end{eqnarray*}}
\def\bi{\begin{itemize}}
\def\ei{\end{itemize}}

\begin{document}
 \title{Density Dependent Parametrization Models: Formalism and Applications}
 \author{S.S. Avancini}
\affiliation{Depto de F\'{\i}sica - CFM - Universidade Federal de Santa
Catarina  Florian\'opolis - SC - CP. 476 - CEP 88.040 - 900 - Brazil}
\author{D.P. Menezes}
\affiliation{Depto de F\'{\i}sica - CFM - Universidade Federal de Santa
Catarina  Florian\'opolis - SC - CP. 476 - CEP 88.040 - 900 - Brazil}
\affiliation{School of Physics - University of Sydney - NSW 2006 - Australia}
 \begin{abstract}
In this work we derive a formalism to incorporate asymmetry and temperature
effects in the Brown-Rho (BR) scaled lagrangian model in a mean field theory.
The lagrangian density discussed in this work requires less parameters than 
the usual models with density dependent couplings. We also present the 
formalism with the inclusion of the eight lightest baryons, two lightest
leptons, $\beta$ equilibrium and charge neutrality in order to apply the BR 
scaled model to the study of neutron stars. The results are again compared with
the ones obtained from another density dependent parametrization model. The 
role played by the rearrangement term at T=0 for nuclear or neutron star 
matter and at finite temperature is investigated. The BR scaled model is shown
to be a good tool in studies involving density dependent effective masses and
in astrophysics applications.
\end{abstract}
 \maketitle
 \vspace{0.50cm}
PACS number(s):21.65.+f, 24.10.Jv,26.60.+c,21.30.Fe
\vspace{0.50cm}
 \section{Introduction}
 The study of nuclear matter properties at high densities and 
at finite temperature has become an important problem since a
large variety of data, where matter is being tested at extreme
conditions of density, pressure and non-zero temperature, are
becoming available in the modern experimental facilities which are
already operational. Moreover, the constitution of the interior of neutron and
protoneutron stars is also a  problem which is receiving much attention in the
recent literature. The crust of the stars, where density is relatively low, can
be adequately described by hadronic models. The correct calculation of the
star properties as the radius and the mass depend on the accuracy of the
equation of state (EOS) used to describe its matter.
 We have checked that different models, either with constant or with density
dependent meson-nucleon couplings present different features at subnuclear
densities of nuclear asymmetric matter by comparing the regions of uniform
unstable matter \cite{chemical}. The parametrizations
of these models generally take into account saturation properties of nuclear
matter and properties of stable nuclei. Extensions of the models for very
asymmetric nuclear matter or to finite temperatures show different behaviors.
 
Another topic of great interest is the in-medium modification of vector meson 
properties. It is well known that the hadron masses are much larger than the
sum of its constituents. One possible explanation for the large masses is that
they may be generated dynamically \cite{wilczek}. Spontaneous breaking of 
chiral symmetry is also related with the hadron masses. At high temperature
and/or dense matter this symmetry is expected to be at least partially 
restored, which modifies the hadron masses and the decay widths 
\cite{hatsuda85},\cite{br}. Experimental signature of the in-medium 
modifications of the vector mesons have been found very recently 
\cite{taps,naruki}. In \cite{taps} the in medium modifications of the $\omega$ 
meson were
investigated in photoproduction experiments and its mass was found to be 
lowered. In \cite{naruki} the vector masses were verified to decrease by
ten percent in medium.  Other experimental results were also reported in 
the same direction \cite{others} and still some experiments have been proposed
to detect in-medium modifications in a near future \cite{proposed}.
 In order to take into account the in-medium modifications of the hadrons,
density dependent relativistic models are certainly more useful than 
models with fixed parameters. In 1991 Brown and Rho
(BR) \cite{br} proposed an in-medium scaling law for the masses
and coupling constants for effective chiral lagrangians. Our
proposal here is to study hadronic properties under extreme
conditions using lagrangians which incorporate BR scaling since
this has been successfully applied to describe meson properties.
As argued in the literature the BR scaling describes the behavior
of  the light mesons in extreme conditions very accurately. For
example the enhancement of dilepton production observed in heavy
ion collisions (S+Au) in the CERES collaboration and (S+W) in the
HELIOS-3 is most economically and beautifully described by a
chiral lagrangian with BR-scaled meson masses \cite{fri,Song}. The
strategy is to assume that the in-medium effective lagrangian has
the same structure as in free space accordingly to the QCD
constraints but with parameters which are modified in the medium. So,
using this approach we obtain an effective theory with
density dependent parameters including many-body correlations in
a practical framework. 
It has also been shown that it is
possible to obtain a relation between the effective parameters of
chiral lagrangians in medium and Landau Fermi Liquid parameters\cite{bengt}.
So, a link between the effective theory of QCD at mean field level
and the many-body theory of nuclear matter is achieved. Using this
reasoning  the authors in \cite{Song} proposed an effective
Lagrangian whose parameters scale in nuclear medium according to
the Brown-Rho (BR) scaling.

We have already shown that one of these density dependent models,
to which we refer as TW model \cite{tw,gaitanos},
originally derived at $T=0$ can be extrapolated to finite temperatures once the
thermodynamical consistency remains unaltered \cite{typel,typel2}.
The important range of temperature which is discussed lies between $10$ and
$150$ MeV since the liquid-gas phase transition takes place around $10$ MeV,
the phase transition from hadronic to quarkionic matter around 150 MeV
and the relevant temperatures in the cooling of a protoneutron star after a
supernova explosion takes place go up to approximately 40 MeV \cite{trapping}.
In this work we discuss another possible application of the
formalism we have derived in \cite{typel,typel2} in order to
incorporate temperature effects in the study of lagrangians with BR scaling. 
 
It is well known that the same relativistic models used in order to explain 
data coming from heavy ion collisions at finite temperature,  
with appropriate parameter sets, also provide EOS
which can be used in the solution of the Tolman-Oppenheimer-Volkoff
differential equations \cite{tov} for the calculation of stellar properties
such as mass, radius and central energy density. Recent measurements of
gravitational redshift of spectral lines provided direct constraints on the 
mass-to-radius ratio \cite{cottam,sanwal}. In this second case, however, the 
interpretation of the absorption features as atomic transition lines is 
controversial \cite{bignami,xu03}. 
In recent works \cite{qmc,deltas} we have checked that there are relativistic
models which can be accommodated within these constraints. Hence, astrophysical
observations can help in the choice of appropriate models to describe hadronic
matter.
 The most common relativistic model used in the description of hadronic matter
is the non linear Walecka model \cite{bb}. When applied in nuclear
astrophysics, this model is normally extended with the inclusion of hyperons,
which are expected to appear at high densities.
It was shown in \cite{alex} that a low effective mass at saturation density
makes the model inappropriate once hyperons are included. The inclusion of
hyperons makes the scalar meson interaction stronger and hence the proton and
neutron effective masses decrease more rapidly with density, acquiring a 
negative value. As a test of the BR scaled model, we also extend it to 
incorporate the eight lightest baryons and enforce $\beta$ equilibrium plus 
charge neutrality by accommodating the two lightest leptons as well.
The same extension is done within the TW model so that both density dependent
models can be compared.
 
Hence, the present work is organised as follows: in section II the formalism
for the BR scaled model and its extension to finite temperature are presented,
the results are compared with the ones obtained with other relativist models 
and a discussion is included. In section III the BR scaled model is modified
so that hyperons can be incorporated and $\beta$ equilibrium and charge 
neutrality are enforced so that an EOS can be obtained and applied to compact 
stars. In the same section the TW model is also considered so that the results 
from two density dependent models are compared.
In section IV a discussion on the role played by the rearrangement term in 
different models is presented. Finally, in the last section, the conclusions 
are drawn.

\section{Formalism - Extension to finite temperature}
  
In its simplest version the Lagrangian density 
reads \cite{Song,Song1} 
\beqa {\cal L} &=&\bar \psi\left[
\gamma_\mu\left(i\partial^{\mu}- g^{\ast}_v(\rho) ~\omega^{\mu}
- g^{\ast}_{\rho}(\rho) \vec{\tau} \cdot \vec{\rho}^\mu
\right)
-M^{\ast}(\rho) + g_s^* (\rho) \phi \right]\psi \nonumber \\
&&+\frac{1}{2}\left(\partial_{\mu}\phi\partial^{\mu}\phi -m_s^{\ast
2}(\rho) \phi^2\right)
-\frac{1}{4}\Omega_{\mu\nu}\Omega^{\mu\nu}+\frac{1}{2}
m_v^{\ast 2}(\rho) \omega_{\mu}\omega^{\mu} \nonumber \\
&&-\frac{1}{4}\vec\rho_{\mu\nu}\cdot\vec \rho^{\mu\nu}+\frac{1}{2}
{m^*_\rho}^2 \vec\rho_{\mu}\cdot \vec \rho^{\mu},
\label{brlag} \;.
\eeqa
where in the notation of \cite{Song1} $\psi$ is the nucleon field,
$\omega_\mu$ the isoscalar vector field, $\phi$ an isoscalar scalar field,
$\Omega_{\mu\nu}=\partial_{\mu}\omega_{\nu}-\partial_{\nu}\omega_{\mu}$,
$\vec{\rho^\mu}$ is the vector isovector field,
$\vec \rho_{\mu\nu}=\partial_{\mu}\vec \rho_{\nu}-\partial_{\nu} \vec
\rho_{\mu} - g^{\ast}_\rho (\vec \rho_\mu \times \vec \rho_\nu)$,
$\vec \tau$ is the isospin operator and the masses with asterisk are BR-scaled
as introduced in \cite{br}:
\beq
\frac{M^*}{M}=\frac{m_s^*}{m_s}=\frac{m_v^*}{m_v}=\frac{m_\rho^*}{m_\rho}=
\Phi(\rho).
\label{br}
\eeq
The scalings of the vector coupling constants are given by
\beq
\frac{g_s^*}{g_s}=\frac{1}{1+x \rho/\rho_0},~~
\frac{g_v^*}{g_v}=\frac{1}{1+z \rho/\rho_0},~~
\frac{g_\rho^*}{g_\rho}=\frac{1}{1+w \rho/\rho_0},
\eeq
where $\rho_0$ is the nuclear saturation density.
In the original papers \cite{Song,rho}, $g_s^*$ was simply taken constant and
equal to $g_s$ and hence, not density dependent and, with this simple choice
a good description of the ground state was obtained. Moreover,
\beq
\Phi(\rho)=\frac{1}{1+y \rho/\rho_0},
\label{phirho}
\eeq
with $y=0.28$ in such a way that $\Phi(\rho_0)=0.78$ \cite{Song} and 
$z$ and $w$ were taken equal or 
tly greater that $y$. 
 In this work we propose another possible parametrization, based on the
works \cite{hatsuda95,naruki}, where 
\beq
\Phi(\rho)=1-y\frac{\rho}{\rho_0},
\label{phirho2}
\eeq
with $y=0.1$ in such a way that a decrease of the meson masses in medium at
the saturation point is ten percent, as found experimentally \cite{naruki}.
 It is also worth mentioning that, as far as we know this is the first work 
where asymmetry is taken into account
by the appropriate inclusion of the $\rho$ meson in a BR scaled model.
One would bear in mind that we are using the letter $\rho$ for both the meson
and for the total baryonic density. As those are common definitions in 
relativistic models we do not believe it may cause any confusion. 
  One can see that the
Lagrangian in (\ref{brlag}) is of the form of a Walecka-type
Lagrangian and all the finite temperature formalism that we have
developed for the density dependent hadron field theory (DDHFT)
\cite{typel,typel2}  can be immediately applied to these
lagrangians. The thermodynamics of effective lagrangians with BR
scaling has been studied in \cite{rho} for zero temperature. The
study of the validity of the BR scaling hypothesis for the
non-zero temperature case is demonstrated in what follows.
 \indent From the Euler-Lagrange equations we obtain the field equations of
motion in the mean field appro\-xi\-ma\-tion for infinite matter, where the
meson fields are replaced by their expectation values.
In this approximation, the expectation
value of the $\sigma$, $\omega$ and $\rho$ meson fields are called $\phi_0$,
$V_0$ and $b_0$ respectively. The coupled equations read
\begin{eqnarray}
{m_s^*}^2\phi_0 - g_s^* \rho_s &=&0, \label{phi} \\
{m_v^*}^2 V_0 - g_v^* \rho&=&0, \label{V0}\\
{m_\rho^*}^2 b_0 -\frac{g_\rho^*}{2} \rho_3&=&0, \label{b0}\\
\left[ i\gamma^{\mu}\partial_\mu -\gamma_0\left( g_v^*V_0+
g^*_{\rho} \tau_3 b_0 + {\Sigma}^{R}_0 \right)
-(M^{\ast}- g_s^* \phi_0) \right] \psi&=&0, \label{dirac}
\end{eqnarray}
where the rearrangement term $\Sigma^{R}_0$ is given by
\begin{equation}
\Sigma^{R}_0=-m_v^* V_0^2 \frac{\partial m_v^*}{\partial \rho}
+ \rho V_0 \frac{\partial g_v^*}{\partial \rho}
- m_\rho^* b_0^2 \frac{\partial m_\rho^*}{\partial \rho}
+ \rho_3 \frac{b_0}{2} \frac{\partial g_{\rho^*}}{\partial \rho}
+ m_s^* \phi_0^2  \frac{\partial m_s^*}{\partial \rho}
+\rho_s  \frac{\partial M^*}{\partial \rho}
-\rho_s \phi_0 \frac{\partial g_s^*}{\partial \rho}
\label{rearr}
\end{equation}
and the scalar and baryonic densities are defined as
\begin{eqnarray}
\rho_s&=&\langle\bar \psi \psi\rangle, \\
\rho&=&\langle\bar \psi \gamma^0 \psi\rangle, \\
\rho_3&=&2 \langle\bar \psi \gamma^0 \tau_3\psi\rangle .
\end{eqnarray}
 Notice that if the original parametrization for the BR-scaled model is used, 
$g_s^*=g_s$ is a constant and the last term of the rearrangement vanishes.
 In the following discussion we consider nuclear matter in the the mean-field
approximation. Due to translational and
rotational invariance the lagrangian density reduces to
\begin{eqnarray}
{\cal L}_{MFT}&=&\bar \psi\left[ i\gamma_\mu \partial^{\mu}
-\gamma_0  g_v^* V_0 - \gamma_0 g^*_{\rho} \tau_3 b_0
-(M^* - g_s^* \phi_0)\right]\psi \nonumber \\
&&
-\frac{1}{2} {m_s^*}^2 \phi_0^2 + \frac{1}{2} {m_v^*}^2 V_{0}^2
+\frac{1}{2} {m^*_\rho}^2 b_{0}^{2} \label{lag2},
\end{eqnarray}
where $\tau_3=\pm 1/2$ for protons and neutrons respectively.
 The conserved energy-momentum tensor can be derived in the usual
fashion \cite{bb}:
 \begin{equation}
{\cal T}_{MFT}^{\mu\nu}=\bar \psi i\gamma^\mu \partial^{\nu}\psi
+ g^{\mu\nu}\left[ \frac{1}{2} {m_s^*}^2 \phi_0^2 - \frac{1}{2} {m_v^*}^2
V_{0}^2
-\frac{1}{2} {m_\rho^*}^2 b_{0}^{2}
+ \bar\psi \gamma_0 \Sigma^{R}_0 \psi \right]  \label{tem} .
\end{equation}
Note that the rearrangement term included above and defined in
eq.(\ref{rearr}) assures the
energy-momentum conservation, i.e., $\partial_\mu{\cal T}^{\mu\nu}=0$. From
the energy-momentum tensor one easily obtains the
hamiltonian operator:
\begin{eqnarray}
 {\cal H}_{MFT}=\int d^3 x ~{\cal T}_{MFT}^{00}&=&
 \int d^3 x ~ \psi^{\dagger}\left(-i\vec\alpha \cdot \nabla
 +\beta m_L^\ast + g_v^* V_0 + g_{\rho}^* \tau_3 b_0
+ \Sigma^R_0 \right) \psi  \nonumber \\
 &&+V\left(\frac{1}{2} {m_s^*}^2 \phi_0^2
 - \frac{1}{2} {m_v^*}^2 V_{0}^2 -\frac{1}{2} {m_\rho^*}^2 b_{0}^{2}
\right), \label{hamil}
\end{eqnarray}
where 
\begin{equation}
m_L^{\ast}=M^*- g_s^* \phi_0
\label{landau1}
\end{equation}
is identified as the effective nucleon Landau 
mass and $V$ is the volume of the system. Notice that the energy density does 
not carry the rearrangement term because it cancels out in a mean field
approximation:
 \begin{equation}
{\cal E}= 2 \sum_{i=p,n} \int \frac{d^3p}{(2\pi)^3}
\sqrt{{\mathbf p}^2+{m^*_L}^2} \left(f_{i+}+f_{i-}\right)
 +\frac{{m_s^*}^2}{2} \phi_0^2+\frac{{m_v^*}^2}{2} V_0^2
+\frac{m_{\rho}^2}{2} b_0^2,\label{enerd}
\end{equation}
where $f_{i+}$ and $f_{i-}$ are the distribution functions for particles and
anti-particles respectively and are calculated next.
 Following the notation in \cite{mp}, the thermodynamic potential
can be written as
\begin{equation}
\Omega= {\cal E} -T {\cal S} - \mu_p \rho_p - \mu_n \rho_n,
\label{Omega1}
\end{equation}
where ${\cal S}$ is the entropy of a classical Fermi gas, $T$ is the
temperature,
$\mu_p$ ($\mu_n$) is the proton (neutron) chemical potential and
$\rho_p$ and $\rho_n$ are respectively the proton and neutron densities,
calculated in such a way that $\rho=\rho_p+\rho_n$. We have
\begin{equation}
\rho_i=2 \int\frac{d^3p}{(2\pi)^3}(f_{i+}-f_{i-}), \quad i=p,n\; ,
\label{rhoi}
\end{equation}
where the distribution functions $f_{i+}$ and $f_{i-}$ for particles and
anti-particles have to be derived in order to make the thermodynamic
potential stationary for a system in equili\-bri\-um. After straightforward
substitutions, eq.(\ref{Omega1}) becomes
 \begin{eqnarray}
\Omega&=&
2 \sum_{i=p,n} \int \frac{d^3p}{(2 \pi)^3} \sqrt{{\mathbf p}^2+{m^*_L}^2}
(f_{i+} + f_{i-})
+\frac{{m_s^*}^2}{2} \phi_0^2+\frac{{m_v^*}^2}{2} V_0^2
+\frac{{m_\rho^*}^2}{2} b_0^2 \nonumber\\
&&+2 T \sum_{i=p,n} \int \frac{d^3p}{(2 \pi)^3} \left(
f_{i+} \ln \left(\frac{f_{i+}}{1-f_{i+}}\right) + \ln ({1-f_{i+}}) +
f_{i-} \ln \left(\frac{f_{i-}}{1-f_{i-}}\right) + \ln ({1-f_{i-}}) \right)
\nonumber \\
&&-2 \sum_{i=p,n} \int \frac{d^3p}{(2 \pi)^3}\:\mu_i (f_{i+}-f_{i-}).
\label{Omega2}
\end{eqnarray}
For a complete demonstration of the above shown expressions
obtained in a Thomas-Fermi approximation for the non-linear Walecka model,
please refer to \cite{mp}.
At this point, eq.(\ref{Omega2}) is minimized in terms of the distribution
functions for fixed meson fields, i.e.,
\begin{equation}
\left. \frac{\partial \Omega}{\partial f_{i+}}
\right |_ {{f_{i-},f_{j\pm}, \phi_0,V_0,b_0}} =0 \quad i \ne j.
\end{equation}
 For the particle distribution function, the above calculation yields
\begin{equation}
E^{\ast}({\mathbf p}) + \Sigma^R_0 - \mu_i + g_v^* V_0 + \frac{g_\rho^*}{2} b_0
= -T\: \ln \left(\frac{f_{i+}}{1-f_{i+}}\right),
\end{equation}
where
$E^{\ast}({\mathbf p})=\sqrt{{\mathbf p}^2+{m_L^*}^2}$.
A similar equation, with a sign difference is obtained for the anti-particle
distribution function. It is important to point out that the fields $\phi_0$, 
$V_0$ and $b_0$ depend on the distribution function which appear in the 
definition of $\rho_s$, $\rho$ and $\rho_3$ and hence, the whole calculation 
is performed self-consistently. The effective chemical potentials are then 
defined as
 \begin{eqnarray}
\nu_p&=&\mu_p- g_v^* V_0 -\frac{g_\rho^*}{2} b_0 -
\Sigma^{R}_0, \nonumber \\
\nu_n&=&\mu_n- g_v^* V_0 +\frac{g_\rho^*}{2} b_0 -
\Sigma^{R}_0 \label{efchem}
\end{eqnarray}
and the following equations for the distribution functions can be written:
\begin{equation}
f_{i\pm}=
\frac{1}{1+\exp[(E^{\ast}({\mathbf p}) \mp\nu_i)/T]}\;,
\quad i=p,n. \label{disfun}
\end{equation}
In the above calculation we have used
\[
\rho_s= 2 \sum_{i=p,n}
\int \frac{d^3p}{(2\pi)^3}
\frac{m^*_L}{E^{\ast}({\mathbf p})}\left(f_{i+}+f_{i-}\right),
\]
and $\rho_3=\rho_p-\rho_n$.
 Within the Thomas-Fermi approach the pressure becomes
\begin{eqnarray}
P&=&\frac{1}{3 \pi^2} \sum_{i=p,n}
\int dp \frac{{\mathbf p}^4}{\sqrt{{\mathbf p}^2+{M^*}^2}} \left( f_{i+} +
f_{i-}\right)\\
&&-\frac{{m_s^*}^2}{2} \phi_0^2 +\frac{{m_v^*}^2}{2} V_0^2
+\frac{{m_\rho^*}^2}{2} b_0^2 + \Sigma^R_0 \rho.
\label{pressure}
\end{eqnarray}
 It is worth mentioning that the thermodynamical consistency
which requires the equality of the pressure calculated from the
thermodynamical definition and from the energy-momentum tensor,
discussed in \cite{typel2}, is also obeyed by the temperature
dependent Brown-Rho scaled model.

 \subsection{Discussions on the BR-scaled model for nuclear matter}

 At this point, the parameters used in the BR method have to be fixed. 
Through out this paper the nucleon mass will be $M=939$ MeV, the $\omega$
meson mass $m_v=783$ MeV and the $\rho$ meson mass $m_\rho=763$ MeV.
Three different sets are used in \cite{rho}. In what follows we use the 
parameter set called S3 in \cite{rho}
and define another one which we call MA, whose bulk properties
are more similar to the NL3 \cite{nl3} parameter set but with a larger
effective mass at nuclear saturation density.
While S3 is a parametrization for the original BR scaled model given by
equation (\ref{phirho}), MA is a parametrization for the new scaling, 
given in equation (\ref{phirho2}).
In Table \ref{tab1} we show the S3 and MA constants and in Table \ref{tab2}
we display the nuclear matter bulk properties described by the different 
models used in this work. It is important to point out that 
the value for the saturation density (vide * in Table II) was not given in 
\cite{Song}. For the
saturation density value shown in Table \ref{tab1}, the compressibility is 
slightly different from what is stated in \cite{Song} (260 MeV). This is 
probably a consequence of the fact that the authors in \cite{Song} have not 
included the rearrangement term in their calculations. Notice that we 
distinguish the $M^*$ from 
the $m^*_L$ values. In the BR scaled models it is the Landau mass that should 
be identified with the nucleon effective mass determined by the QCD sum rule
\cite{Song} and its value should lie in between 0.55 M and 0.75 M.
  \begin{table}[ht]
\caption{ Parameter sets for the Lagrangian (\ref{brlag})}
\centering
\begin{tabular}{|c|c|c|c|c|c|c|c|c|c|c|}
\hline
Set  &  $m_s$ & $g_s$ &  $g_v$ & $g_{\rho}$ & $x$ & $z$ & $w$ \\
\hline
S3  & 700 & 5.30 & 15.2   & 7.97  & -    & 0.31  & 0.31 \\
MA  & 500 & 7.05 & 12.006 & 8.761 & 0.37 & 0.15  & 0.15   \\
\hline
\end{tabular}
\label{tab1}
\end{table}
  \begin{table}[ht]
\caption{Nuclear matter properties}
\centering
\begin{tabular}{|c|c|c|c|c|c|c|c|c|c|c|}
\hline
   &  NL3 \cite{nl3}  & TM1 \cite{tm1} & GL \cite{Glen00} &
 TW \cite{tw} & S3 \cite{rho} & MA\\
\hline
$B/A$ (MeV) & 16.3 & 16.3 & 15.95 & 16.3 & 16.1 & 16.3\\
\hline
$\rho_0$ (fm$^{-3}$) & 0.148 & 0.145 & 0.145 & 0.148 & 0.155(*) & 0.148\\ 
\hline
$K$ (MeV) & 272 & 281 & 285 & 240 & 269 & 258\\
\hline
${\cal E}_{sym.}$ (MeV)  & 37.4 & 36.9 & 36.8 & 32.0 & 32.0 & 32.0\\
\hline
$M^*/M$ & 0.60 & 0.63 & 0.77 & 0.56 & 0.78  & 0.9\\
\hline
$m^*_L/M$ & - & - & - & - & 0.68  & 0.748\\
\hline
\end{tabular}
\label{tab2}
\end{table}
 
\begin{figure}
\begin{center}
\includegraphics[width=10.cm,angle=0]{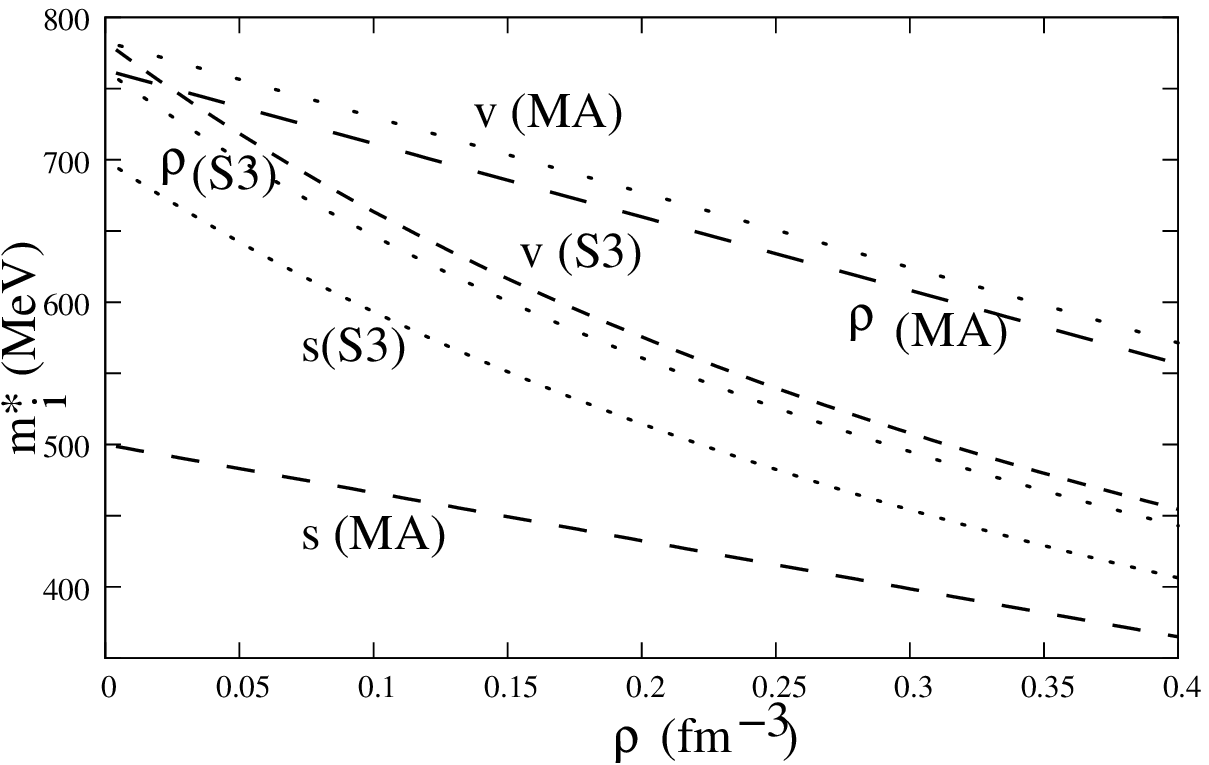}
\caption{Meson masses $m^*_i$,$i=s,v,\rho$ versus the baryonic density.} 
\label{effmass}
\end{center}
\end{figure}

\begin{figure}
\begin{center}
\includegraphics[width=10.cm,angle=0]{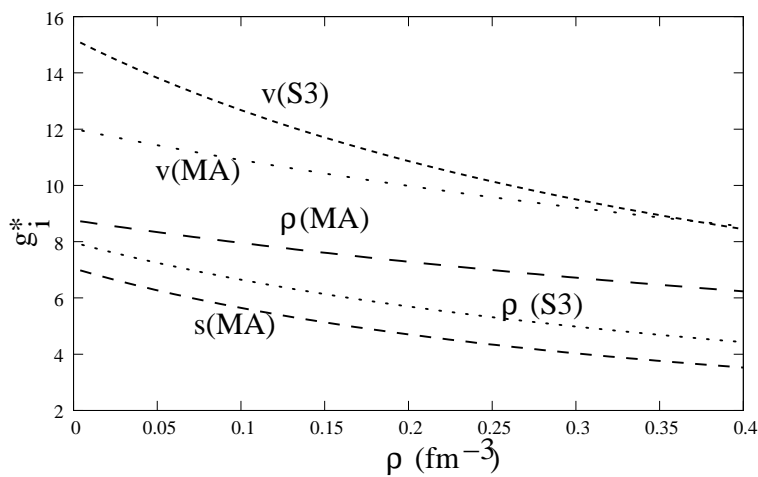}
\caption{Meson to baryon couplings ${g^*_i}^2$, $i=s,v,\rho$
in terms of the density.}
\label{effctes}
\end{center}

\end{figure}

\begin{figure}
\includegraphics[width=10.cm,angle=0]{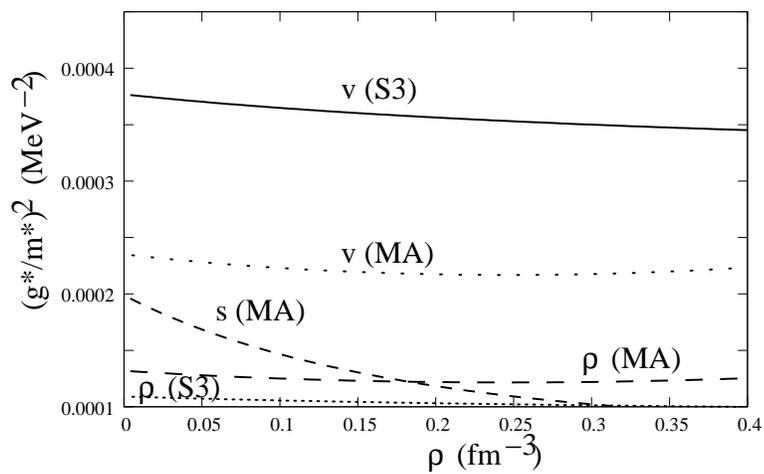}
\caption{Ratios ${g^*_i}^2/{m^*_i}^2$, $i=s,v,\rho$ as function of the 
density.} 
\label{efftudo}
\end{figure}

In figure \ref{effmass} we plot the dependence of the meson
masses with the density for the S3 and MA parameter sets. This is an important
consequence of this model, since the reduction of the meson masses in medium
is an expected result \cite{hatsuda95}. As stated in the Introduction, this 
behavior is related with the
restoration of the chiral symmetry and experiments with the spectrometer HADES
at GSI will also be measuring this effect soon. If the S3 parametrization is 
used, the effective masses of all mesons decrease by 22$\%$ up to the 
saturation density while if the MA parametrization is used, the decrease 
is forced to be just 10$\%$, as found in \cite{naruki}.
 In figure \ref{effctes} we display the behavior of the coupling
constants. Although both $g_v^*$'s are quite different at subsaturation 
densities, they tend to achieve reasonably close values at larger densities. 
$g_{\rho}^*$, 
on the other hand, presents quite a similar behavior in both models and 
$g_s^*$ only changes with density within the MA framework. 
 In figure \ref{efftudo} we show the ratios $(g_i^*/m_i^*)^2$, with 
$i=s,v,\rho$ which are quantities always present in nuclear matter 
calculations. One can see that the ratios are very small, the ratio involving
the $\omega$ meson being the largest in both parametrizations.
One should bear in mind that, as stated earlier, within the original version 
of the BR scaled model, the scalar coupling constant does not vary with the 
density. For the MA set, on the other hand, the ratio $(g_s^*/m_s^*)^2$ tends 
to zero at $\simeq 2 \rho_0$.
 
\begin{figure}
\begin{tabular}{cc}
\includegraphics[width=10.cm,angle=0]{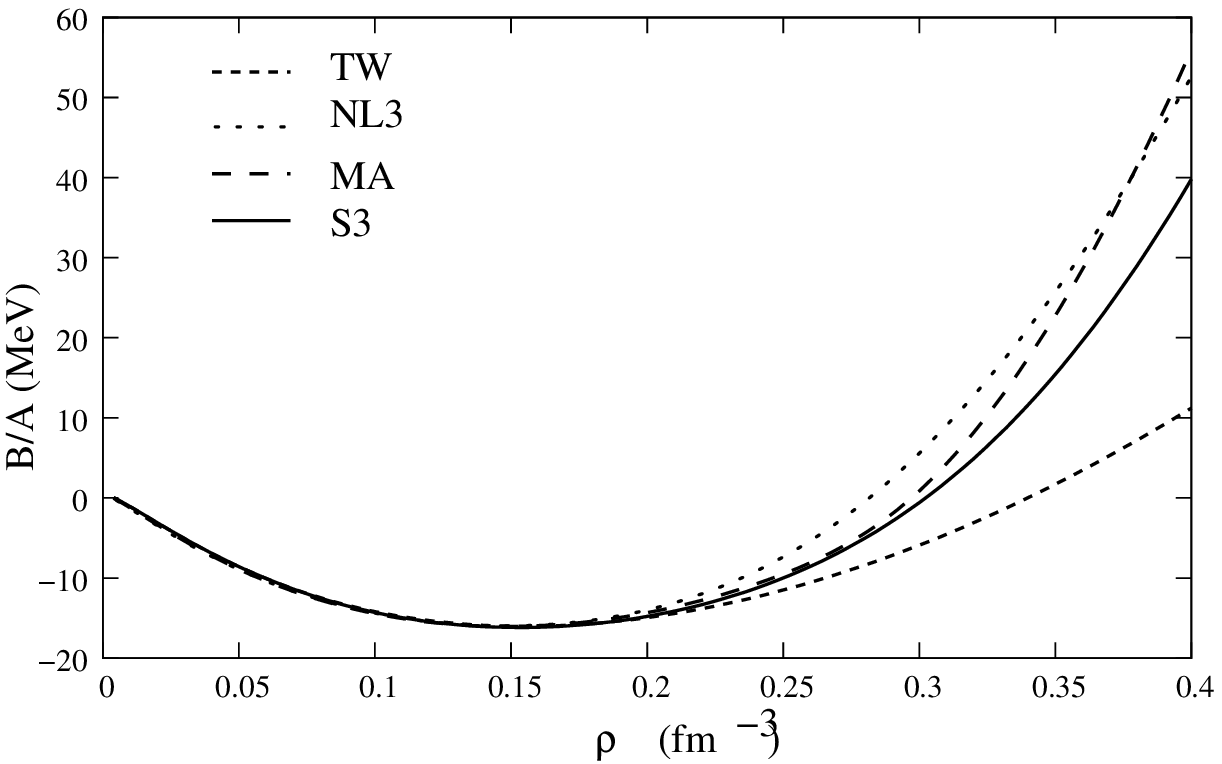}\\
\includegraphics[width=10.cm,angle=0]{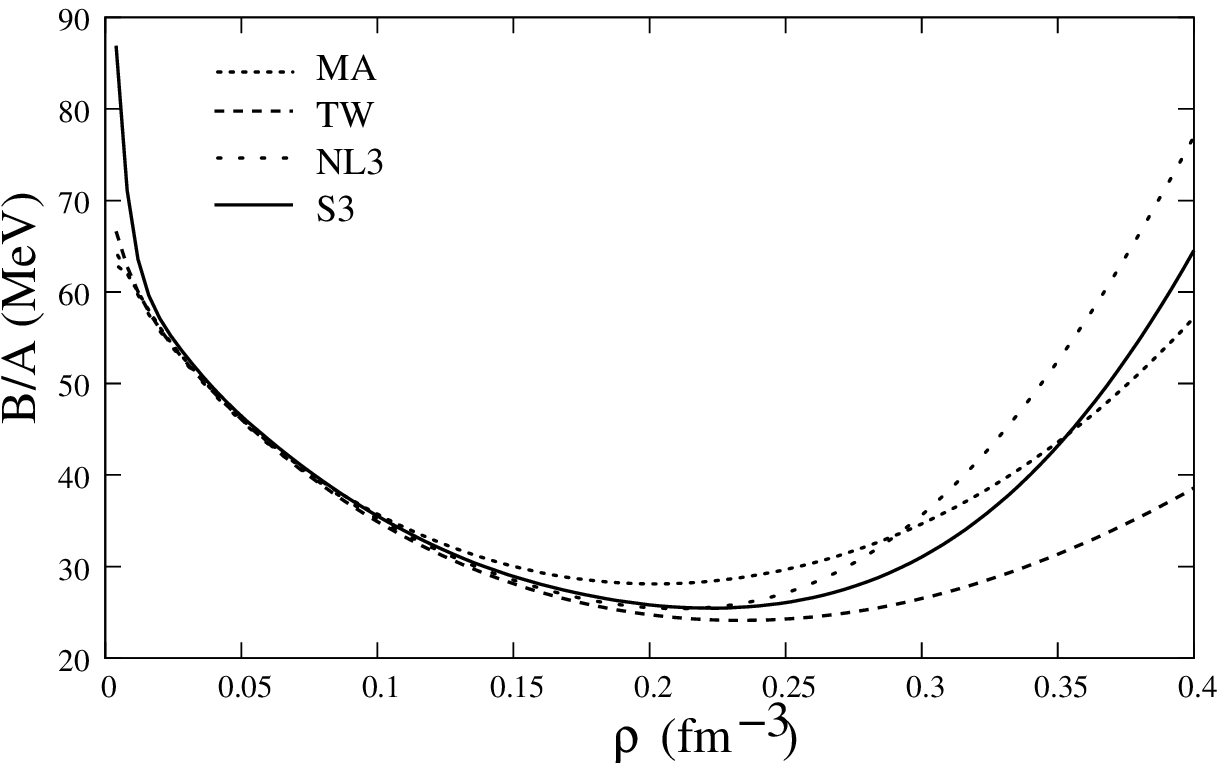}\\
\end{tabular}
\caption{Binding energy in terms of the baryon density for different models
at $T=0$ (top figure) and $T=40$ MeV (bottom figure).} 
\label{fig1}
\end{figure}

In figure \ref{fig1} we show the binding energy in terms of
the baryon density for different models for $T=0$ and $T=40$ MeV.
For the sake of comparison with other models, we have chosen one model with
constant couplings (NL3) and another one with density dependent couplings (TW).
At $T=0$, the TW model is the softest one and the NL3 the hardest, the two
curves obtained with the BR scaled parametrizations interpolating between the
other models.
One can see that the temperature does not alter the softness (hardness) of the
EOS considered. The hardest and the softest ones at $T=0$ remain so at a 
higher temperature.

\begin{figure}
\begin{center}
\includegraphics[width=10.cm,angle=0]{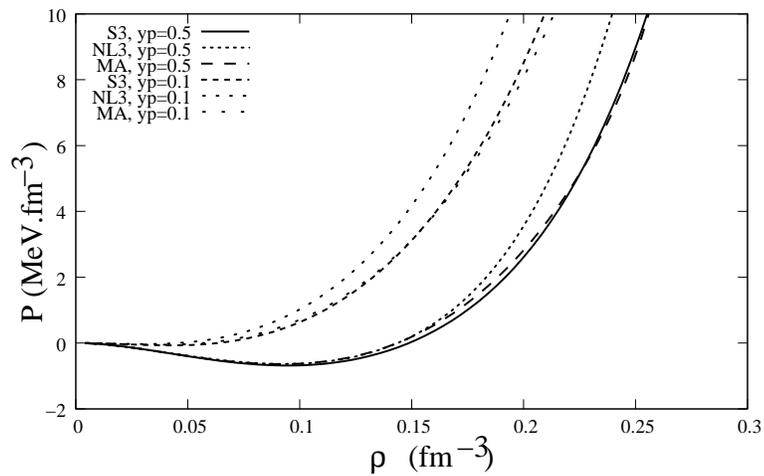}
\caption{Pressure in terms of the baryon density for different models
at $T=0$ and two different proton fractions.} 
\label{fig2}
\end{center}
\end{figure}

In figure \ref{fig2} the pressure versus the baryon density is displayed for
symmetric nuclear matter ($y_p=0.5$) and for very asymmetric matter
($y_p=0.1$), where $y_p=\rho_p/\rho$ is the proton fraction. The isospin is a
quantity which influences the softness (hardness) of the EOS, but one can see 
that the asymmetry seems not to affect the displayed EOSs. 

\begin{figure}
\begin{center}
\includegraphics[width=10.cm,angle=0]{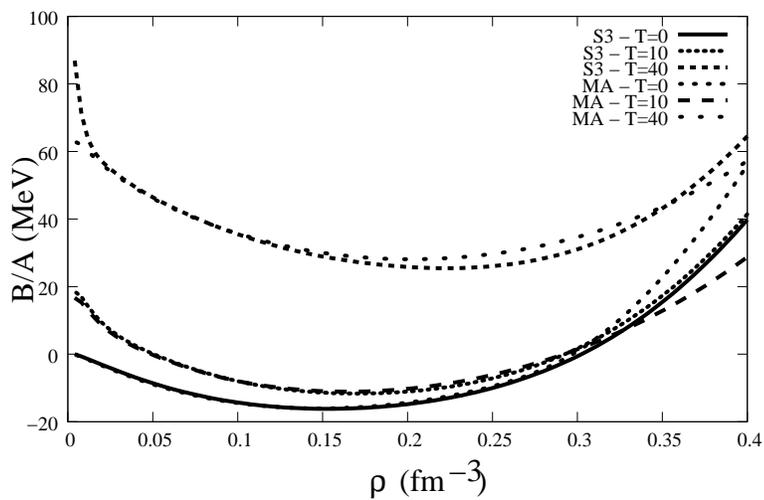}
\caption{Binding energy in terms of the baryon density for S3 and MA and
different temperatures.} 
\label{fig3}
\end{center}
\end{figure}

In figure \ref{fig3} we show how the temperature affects the binding energy of
the S3 and MA models for symmetric matter. The behavior is the same one
encountered in \cite{typel}, i.e., the minimum shifts to higher densities with
the increase of the temperature and moves from a negative to a positive value.
This seems to be a natural consequence of the increase in the temperature of
the system.

\begin{figure}
\begin{center}
\includegraphics[width=10.cm,angle=0]{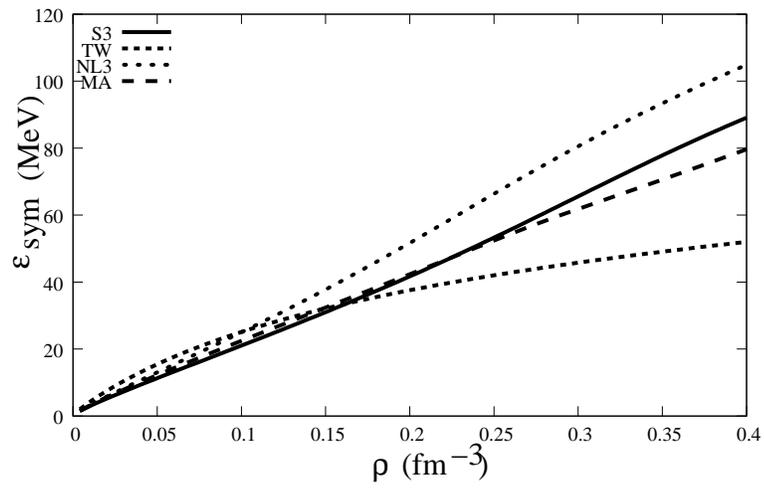}
\caption{Symmetry energy in terms of the baryon density for different models
at $T=0$.} 
\label{fig4}
\end{center}
\end{figure}

 Another quantity of interest is the nuclear bulk symmetry energy discussed
in \cite{lkb}. For symmetric nuclear matter at T=0 it is defined as
\begin{equation}
{\cal E}_{sym}= \frac{P_F^2}{6 E^{\ast}(P_F)}+ \frac{{g_\rho^*}^2}
{8 {m_\rho^*}^2} \rho, \label{esym}
\end{equation}
with $P_F=(1.5 \pi^2\rho)^{1/3}$.
The value and behavior of the symmetry energy at densities larger than nuclear
saturation density are still not well established. This quantity is important
in studies involving neutron skins, radioactive nuclei and neutron stars. In
general, relativistic and non-relativistic models give different predictions
for the symmetry energy. The results of this quantity for different models
are also discussed in the present work and the values at saturation density
are shown in Table \ref{tab2}.
In figure \ref{fig4} we plot the symmetry energy for the BR scaled
models, NL3 and TW. In S3, MA and TW the symmetry energy at saturation density
is the same. The three curves do not cross in the same point because the
saturation densities are not the same (see Table \ref{tab2}). Notice that
although
S3 and MA present a lower symmetry energy at subsaturation densities, they
interpolate between NL3 and TW at larger densities. The value of 32 MeV
that we have chosen for the symmetry energy in order to fix the $g_{\rho}$
coupling is lower than the ones found in most relativistic models
(between 35 and 42 MeV) and approaches the values obtained in non-relativistic
models (between 28 and 38 MeV). 
Notice that the choice of parameters is not arbitrary. 
They are chosen in order to reproduce the nuclear bulk properties of Table 
\ref{tab2}. 
Moreover, they also have to give the correct value of the spin-orbit splitting 
strength. Work in the direction of calculating this quantity in finite nuclei 
is in progress, which may require small changes in the calculated parameter 
sets.

 \section{Application to compact stars}

\noindent From the results shown in the previous section we could see that the 
BR scaled
models show quite a different behavior from the NL3 and TW models
at high densities. In what follows we intend to investigate which are the 
consequences of using the BR scaled model in the description of neutron star
matter. The behavior of the EOS at high densities is responsible for the 
determination of the maximum mass of the star. 
In order to apply the BR scaled density dependent model to compact stellar 
objects, it is important to allow for the inclusion of the eight lightest
baryons (nucleons, $\Lambda$, $\Sigma^0$, $\Sigma^{\pm}$,$\Xi^-$ and $\Xi^0$)
as well as the two lightest leptons ($e^-$ and $\mu$). The baryons have to
be considered since their masses are such that their presence is already 
possible at the neutron stars high densities. The leptons, on the other hand,
play a decisive role in ensuring charge neutrality and $\beta$ equilibrium.
As seen in equation (\ref{efchem}) the rearrangement term which appears
due to the density dependent couplings alters the chemical potentials of the
particles in the system and hence, the $\beta$ equilibrium conditions are
somewhat different as compared with the usual relativistic models. In what 
follows we show the formalism developed for the BR scaled model in neutron
stars and also for the TW model, so that two density dependent parametrization
models can be compared.
 
\subsection{Considering $\beta$ equilibrium within the BR-scaled model}

\noindent For our purposes of testing the BR scaled model at high densities, 
which are
present in neutron stars, we shall restrict ourselves to the $T=0$ case.
Of course, the extension to finite temperature is trivial and can be done 
following the steps of section II. Actually in calculations involving 
protoneutron stars or stars with fixed entropy and trapped neutrinos, the
extension has to be done. 
 Equation (\ref{brlag}) is then modified in order to accommodate these new
particles
 \beqa {\cal L} &=& \sum_B \bar \psi_B \left[
\gamma_\mu\left(i\partial^{\mu}- g^{\ast}_{v B}(\rho) ~\omega^{\mu}
-g^{\ast}_{\rho B}(\rho) \vec{\tau} \cdot \vec{\rho}^\mu
\right)
-M^{\ast}_B (\rho) + g^*_{sB}(\rho) \phi \right]\psi_B \nonumber \\
&&+\frac{1}{2}\left(\partial_{\mu}\phi\partial^{\mu}\phi -m_s^{\ast
2}(\rho) \phi^2\right)
-\frac{1}{4}\Omega_{\mu\nu}\Omega^{\mu\nu}+\frac{1}{2}
m_v^{\ast 2}(\rho) \omega_{\mu}\omega^{\mu} \nonumber \\
&&-\frac{1}{4}\vec\rho_{\mu\nu}\cdot\vec \rho^{\mu\nu}+\frac{1}{2}
{m^*_\rho}^2 \vec\rho_{\mu}\cdot \vec \rho^{\mu}
+ \sum_l \bar \psi_l \left(i \gamma_\mu \partial^{\mu}-
m_l\right)\psi_l
\label{brlag2} \;.
\eeqa
where the meson field operators represent the same mesons as in 
eq.(\ref{brlag}), $\psi_B$ now represents each of the eight baryons,  
$l$ describes the two leptons whose masses are respectively $m_e=0.511$ MeV 
and $m_{\mu}=106.55$ MeV 
and the masses with asterisk are again BR-scaled as in eq. (\ref{br}), the 8
baryons of the octet also obeying the same scaling law, i.e.,
\beq
\frac{M^*_B}{M_B}= \Phi(\rho).
\eeq
The baryon meson couplings are defined as
$g_{sB}^*=x_{s B}~ g_s^*,~~g_{v B}^*=x_{v B}~ g_v^*,~~
g_{\rho B}^*=x_{\rho B}~ g_{\rho}^*$
and $x_{s B}$, $x_{v B}$ and $x_{\rho B}$ are equal to $1$ for the nucleons 
and may have different values for the hyperons.
 Again the meson fields are obtained in the same way as in section II and they
now read:
\begin{eqnarray}
{m_s^*}^2\phi_0 -\sum_B g_{sB}^* \rho_{s B} &=&0, \label{phibr} \\
{m_v^*}^2 V_0 - \sum_B g_{v B}^* \rho_B&=&0, \label{V0br}\\
{m_\rho^*}^2 b_0 -\sum_B g_{\rho B}^*~ \tau_{3B}~ \rho_B&=&0, \label{b0br}\\
\left[ i\gamma^{\mu}\partial_\mu -\gamma_0\left( g_{v B}^* V_0+
g^*_{\rho B}~\tau_{3B}~ b_0 + {\Sigma^{R}_0}_{BR} \right)
-(M^{\ast}_B- g_{sB}^* \phi_0) \right] \psi&=&0, \label{diracbr}
\end{eqnarray}
where the term ${\Sigma^{R}_0}_{BR}$ is now changed and is given by
$$
{\Sigma^{R}_0}_{BR}=-m_v^* V_0^2 \frac{\partial m_v^*}{\partial \rho}
+ \sum_B \rho_B V_0 \frac{\partial g_{v B}^*}{\partial \rho}
- m_\rho^* b_0^2 \frac{\partial m_\rho^*}{\partial \rho}
+ \sum_B \tau_{3B}~ \rho_B~ b_0 \frac{\partial g_{\rho B}^*}{\partial \rho}
$$
\begin{equation}
+ m_s^* \phi_0^2  \frac{\partial m_s^*}{\partial \rho}
+\sum_B \rho_{s B}  \frac{\partial M^*_B}{\partial \rho}
-\rho_{sB} \phi_0 \frac{\partial g_{sB}^*}{\partial \rho},
\label{rearrbr}
\end{equation}
$\tau_{3B}$ is the isospin projection of each baryon and
the scalar and baryonic densities are 
\begin{eqnarray}
\rho_{s B}&=& \frac{1}{\pi^2} \int p^2 dp 
\frac{m^*_{L B}}{E^*_B},\label{rhos}\\
\rho_B&=& \frac{k_{FB}^3}{3 \pi^2}, \label{rhob}
\end{eqnarray}
with
\begin{eqnarray}
m^*_{L B}&=&M^*_B- g_{sB}^* \phi_0, \\
E^*_B&=&\sqrt{{\mathbf p}^2+{m^*_{L B}}^2},\\
\nu_B&=& \mu_B-g_{v B}^* V_0 - g_{\rho B}^*~ \tau_{3B}~ b_0 
- {\Sigma^{R}_0}_{BR} \\
&=& \sqrt{k_{FB}^2 + {m^*_{LB}}^2}
\label{effbr}.
\end{eqnarray}

 The equation of state, necessary for the description of the stellar matter can
now be obtained. The energy density and the pressure density are given 
respectively by
 \beq
{\cal E}=\frac{1}{\pi^2} \sum_B \int_0^{k_{FB}} p^2 dp~ E^*_B + 
\frac{{m^*_s}^2}{2} \phi_0^2 + \frac{{m^*_v}^2}{2} V_0^2
+ \frac{{m^*_{\rho}}^2}{2} b_0^2 
+ \frac{1}{\pi^2} \sum_l \int_0^{k_{Fl}} p^2 dp~ E_l
\eeq
and
\beq
P=\frac{1}{3 \pi^2} \sum_B \int_0^{k_{FB}} \frac{p^4 dp}{E^*_B}  
-\frac{{m^*_s}^2}{2} \phi_0^2 + \frac{{m^*_v}^2}{2} V_0^2
+ \frac{{m^*_{\rho}}^2}{2} b_0^2 
+ (\sum_B \rho_B) {\Sigma^R_0}_{BR}
+ \frac{1}{3 \pi^2} \sum_l \int_0^{k_{Fl}} \frac{p^4 dp}{E_l},
\eeq
where $E_l=\sqrt{{\mathbf p}^2 + m_l^2}$,~$\rho_l=\frac{k_{Fl}^3}{3 \pi^2}$
and $k_{Fl}=\sqrt{\mu_l^2- m_l^2}$, $\mu_l$ being the chemical potential of 
the lepton $l$.
 Notice that, as far as the leptons do not exchange mesons with the baryons nor 
with themselves, they were introduced as free Fermi gases.
The weak interaction between leptons and hadrons is taken
into account through the constraints of charge neutrality and $\beta$ 
equilibrium given respectively by:
\begin{equation}
\sum_B q_B^e \rho_B+ \sum_l q_l^e \rho_l =0,
\label{charge}
\end{equation}
where $q_B^e$ is the electric charge of baryon $B$,
$q_l^e$ is the electric charge of lepton $l$ and
\begin{equation}
\mu_B=\mu_n-q^e_B \mu_e \label{chem}.
\label{beta}
\end{equation}

 \subsection{Considering $\beta$ equilibrium within the TW model}

 In order to make a comparison with the density dependent BR scaled model, 
we next make some considerations about the TW model \cite{tw},
originally derived at $T=0$ and which has also been extrapolated to finite 
temperatures \cite{typel,typel2}. In what follows we write the most important
formulae for the TW model once the lightest baryon octet and the lightest 
leptons are included and charge neutrality and $\beta$ equilibrium are 
enforced. In reference \cite{alemaes} a similar approach was developed and two
different models were discussed. In the first of them the couplings depend on 
the total baryonic density, as done in the BR scaled approach shown in the last
subsection and also next in
the present work. In the second model the couplings depend only on the proton 
plus neutron densities. The authors of reference \cite{alemaes} showed that an
examination of the neutron star properties favored the first model.
 Notice that we next redefine many of the previously defined
quantities, as the baryon effective mass, baryon chemical potentials, etc.
The new equations should not be mixed up with the equations given in the 
previous subsections although we have kept the same notation.

 We start from the lagrangian density
 \beqa {\cal L}&=& \sum_B \bar \psi_B \left[
\gamma_\mu\left(i\partial^{\mu}-\Gamma_{v B} \omega^{\mu}-
\Gamma_{\rho B} \vec{\tau} \cdot \vec{\rho}^\mu \right)
-(M-\Gamma_{s B} \phi)\right]\psi_B \nonumber \\
&&+\frac{1}{2}(\partial_{\mu} \phi \partial^{\mu}\phi
-m_s^2 \phi^2)
-\frac{1}{4}\Omega_{\mu\nu}\Omega^{\mu\nu}+\frac{1}{2}
m_v^2 \omega_{\mu}\omega^{\mu} \nonumber \\
&&-\frac{1}{4}\vec\rho_{\mu\nu}\cdot\vec \rho^{\mu\nu}+\frac{1}{2}
m_\rho^2 \vec\rho_{\mu}\cdot \vec \rho^{\mu}
+ \sum_l \bar \psi_l \left(i \gamma_\mu \partial^{\mu}-
m_l\right)\psi_l,
\label{lagtw}
\eeqa
with all the definitions for the fields given after eq. (\ref{brlag2})
still holding. $\Gamma_{iB}$ and $m_i$ are respectively the couplings of the 
mesons $i=s,v,\rho$  with the hyperons and their masses. 
In this model the set of constants is defined by
$\Gamma_{s B}=x_{s B}~ \Gamma_s,~~\Gamma_{v B}=x_{v B}~ \Gamma_v,~~
\Gamma_{\rho B}=x_{\rho B}~ \Gamma_{\rho}$
and as in subsection III A, 
$x_{s B}$, $x_{v B}$ and $x_{\rho B}$ are equal to $1$ for the nucleons
and can acquire different values for the hyperons. 
$\Gamma_s$, $\Gamma_v$ and $\Gamma_\rho$ are the nucleon-meson 
coupling constants which are adjusted in order to reproduce
some of the nuclear matter bulk properties, using the following 
parametrization:
 \begin{equation}
\Gamma_i(\rho)=\Gamma_i(\rho_{0})f_i(x), \quad i=s,v
\label{paratw1}
\end{equation}
with
\begin{equation}
f_i(x)=a_i \frac{1+b_i(x+d_i)^2}{1+c_i(x+d_i)^2},
\end{equation}
where $x=\rho/\rho_{0}$ and
\begin{equation}
\Gamma_{\rho}(\rho)=\Gamma_{\rho}(\rho_{0}) \exp[-a_{\rho}(x-1)],
\label{paratw2}
\end{equation}
with the values of the parameters $m_j$, $\Gamma_j$, $a_j$, $b_i$, $c_i$ and 
$d_i$, $j=s,v,\rho$  given in \cite{tw}. The nucleon, $\omega$ and 
$\rho$ meson masses are taken as in the BR scaled model. The scalar meson mass
$m_s$ is 500 MeV. Other possibilities
for these parameters are also found in the literature \cite{ditoro}.
The meson and baryon coupled equations for the fields read:
\begin{eqnarray}
m_{s}^2\phi_0 
- \sum_B \Gamma_{sB} ~\rho_{s B} &=&0, \label{phitw} \\
m_{v}^2 V_0 - \sum_B \Gamma_{v B}~ \rho_B&=&0, \label{V0tw}\\
m_\rho^2 b_0 -\sum_B \Gamma_{\rho B}~ \tau_{3B}~ \rho_B&=&0, 
\label{b0tw}\\
\left[ i\gamma^{\mu}\partial_\mu -\gamma_0\left( \Gamma_{v B}
V_0+ \Gamma_{\rho B}~ \tau_{3B}~ b_0 +{{\Sigma}^{R}_0}_{TW} \right)
-M^{\ast}_B \right] \psi&=&0, \label{diractw}
\end{eqnarray}
 where the term ${\Sigma^{R}_0}_{TW}$ is given by
\begin{equation}
{\Sigma^{R}_0}_{TW}=\sum_B \left[ 
\frac{\partial \Gamma_{v B}}{\partial \rho} \rho_B V_0 +
\frac{\partial \Gamma_{\rho B}}{\partial \rho} \tau_{3b}~ \rho_B~ b_0 -
\frac{\partial \Gamma_{s B}}{\partial \rho} \rho_{s B} \phi_0,  
\right]
\label{rearrtw}
\end{equation}
 and the scalar and baryonic densities are defined as
\begin{eqnarray}
\rho_{s B}&=& \frac{1}{\pi^2} \int p^2 dp \frac{M^*_B}{E^*_B},\label{rhostw}\\
\rho_B&=& \frac{K_{FB}^3}{3 \pi^2} \label{rhobtw},
\end{eqnarray}
with
\begin{eqnarray}
M^*_B&=&M_B- \Gamma_{s B} \phi_0, \\
E^*_B&=&\sqrt{{\mathbf p}^2+{M^*_B}^2}.
\end{eqnarray}
Notice that the rearrangement term shown in equation (\ref{rearrtw}) is the 
same one shown
in equation (18) of \cite{alemaes}, once the delta and the strange mesons are
excluded from their calculation.
 The effective chemical potentials are then defined as
\begin{equation}
\nu_B= \mu_B-\Gamma_{v B} V_0 - \Gamma_{\rho B}~ \tau_{3B}~ b_0 
- {\Sigma^{R}_0}_{TW} = \sqrt{k_{FB}^2+{M^*_B}^2}
\label{efftw}.
\end{equation}
 The conditions for $\beta$ equilibrium and charge neutrality are again the 
same ones as in subsection III A, given by eqs.(\ref{charge}) and 
(\ref{beta}). The
final expressions for the energy density and pressure become respectively:
\beq
{\cal E}=\frac{1}{\pi^2} \sum_B \int_0^{k_{FB}} p^2 dp~ E^*_B + 
\frac{m_{s}^2}{2} \phi_0^2 + \frac{m_{v}^2}{2} V_0^2
+ \frac{m_{\rho}^2}{2} b_0^2 
+ \frac{1}{\pi^2} \sum_l \int_0^{k_{Fl}} p^2 dp~ E_l
\eeq
and
\beq
P=\frac{1}{3 \pi^2} \sum_B \int_0^{k_{FB}} \frac{p^4 dp}{E^*_B}  
- \frac{m_{s}^2}{2} \phi_0^2 + \frac{m_{v}^2}{2} V_0^2
+ \frac{m_{\rho}^2}{2} b_0^2 
+ (\sum_B \rho_B) {\Sigma^R_0}_{TW}
+ \frac{1}{3 \pi^2} \sum_l \int_0^{k_{Fl}} \frac{p^4 dp}{E_l}.
\eeq
 \subsection{Discussions on the compact star properties}
 Although NL3 \cite{nl3} and TM1 \cite{tm1} are the most common parametrizations
of the NLWM for nuclear matter and finite nuclei studies,
it is well known that they are just adequate for 
the description of neutron star properties if only protons, neutrons and 
leptons are considered as possible constituents \cite{alex}. The inclusion of
hyperons softens the EOS, but also makes the baryon effective masses decrease
too fast and a good convergence can only be obtained at relatively low
densities. For this reason, whenever hyperons are considered in the present
work, we shall make comparisons with the GL \cite{Glen00} parametrization of 
the NLWM, where the above mentioned problem does not exist.
It is our aim also to verify whether this problem is present in the TW and
BR scaled model.

 At this point the meson-hyperon couplings have to be fixed. Several 
possibilities are discussed in the literature \cite{Glen00,ghosh}. According  
to \cite{gm91,Glen00} the hyperon couplings constrained by the 
binding of the $\Lambda$ hyperon in nuclear matter, hypernuclear levels and 
neutron star masses yields
$x_{s B}=0.7$ and $x_{v B}=x_{\rho B}=0.783$ and the couplings to the 
$\Sigma$ and $\Xi$ are equal to those of the $\Lambda$ hyperon. Another
possibility is to take $x_{s B}=x_{v B}= x_{\rho B}=\sqrt{2/3}$ as in 
\cite{moszk,gl89,ghosh}. This choice is based on quark counting arguments.
The  universal coupling, with $x_{s B}=x_{v B}= x_{\rho B}=1$ has also been
used \cite{glen85}. From \cite{aquino,deltas} it can be verified that the 
compact star properties depend on the choice of these parameters. As our aim in
the present work is to compare results obtained from different models and the
correct choice is still not well established, we have used the simple 
universal coupling in what follows.

\begin{figure}
\begin{tabular}{cc}
\includegraphics[width=7.2cm,angle=0]{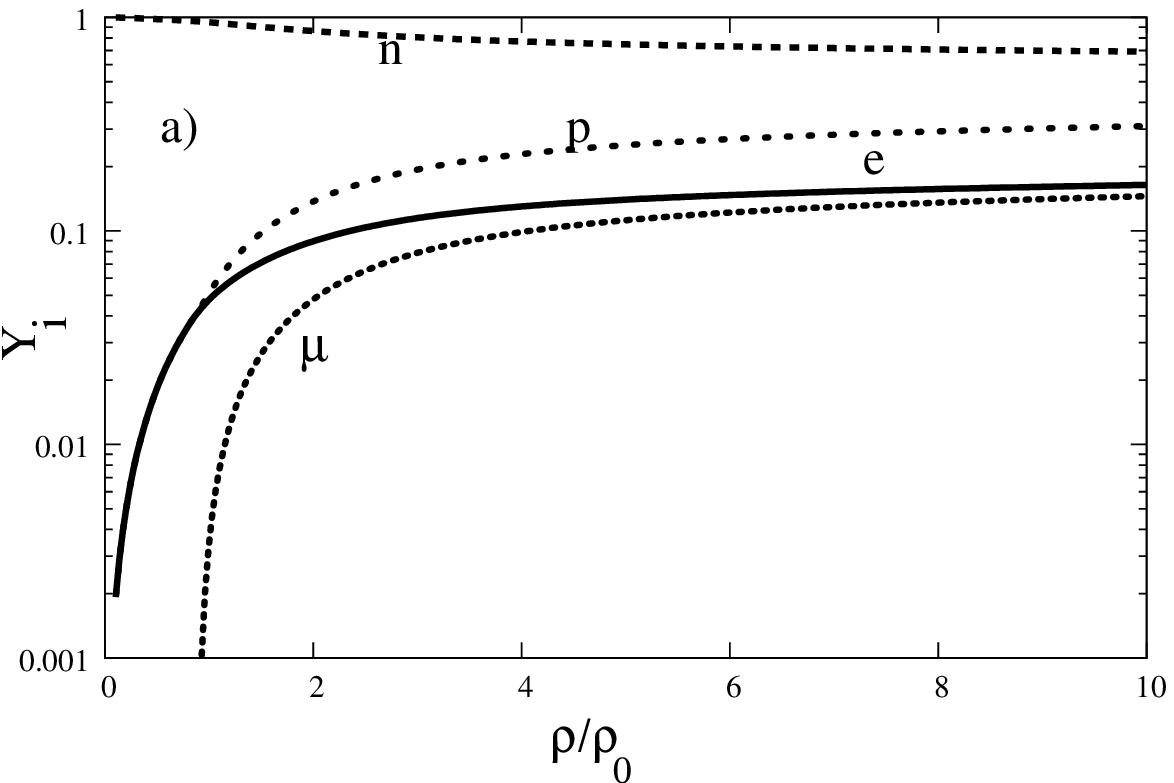} &
\includegraphics[width=7.2cm,angle=0]{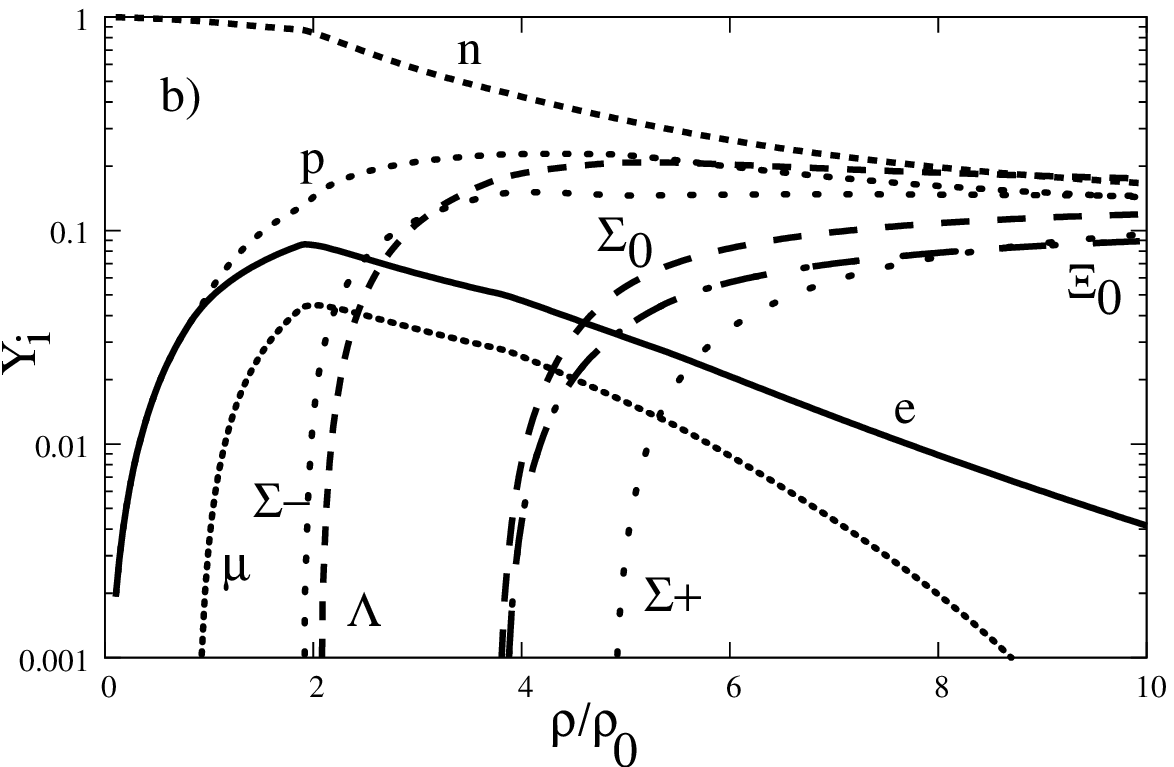} \\
\includegraphics[width=7.2cm,angle=0]{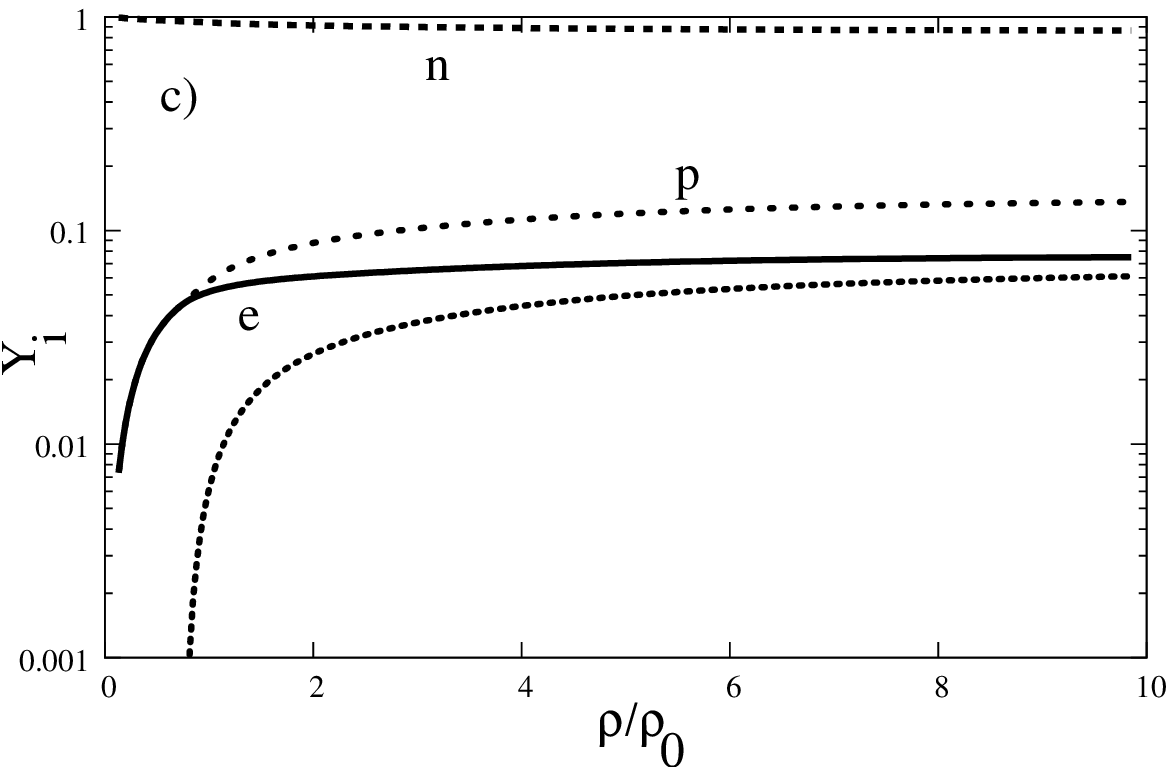} & 
\includegraphics[width=7.2cm,angle=0]{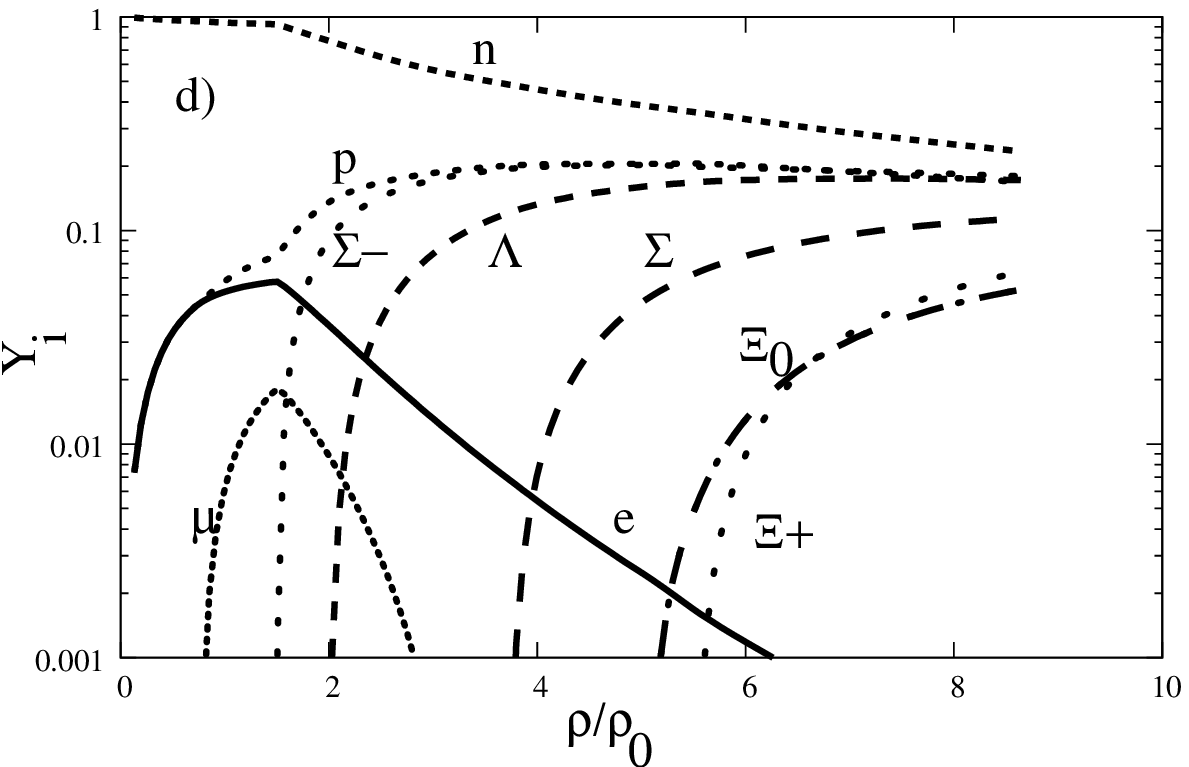}\\
\includegraphics[width=7.2cm,angle=0]{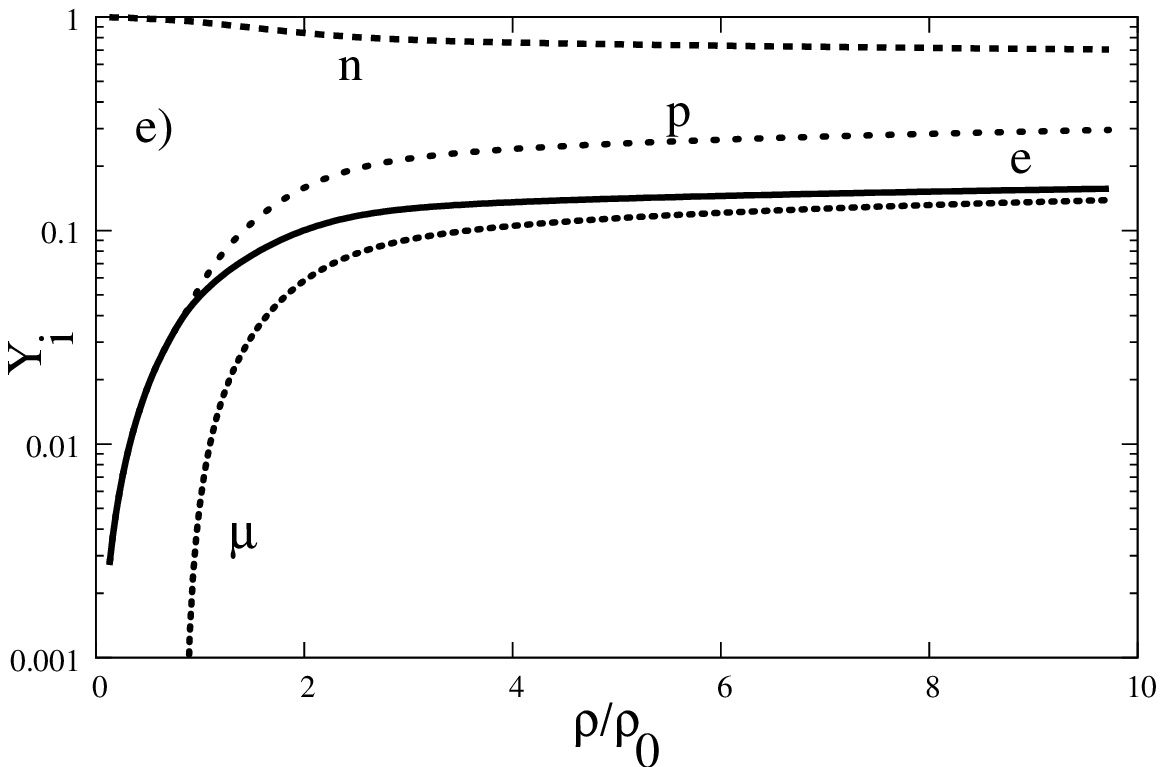} &
\includegraphics[width=7.2cm,angle=0]{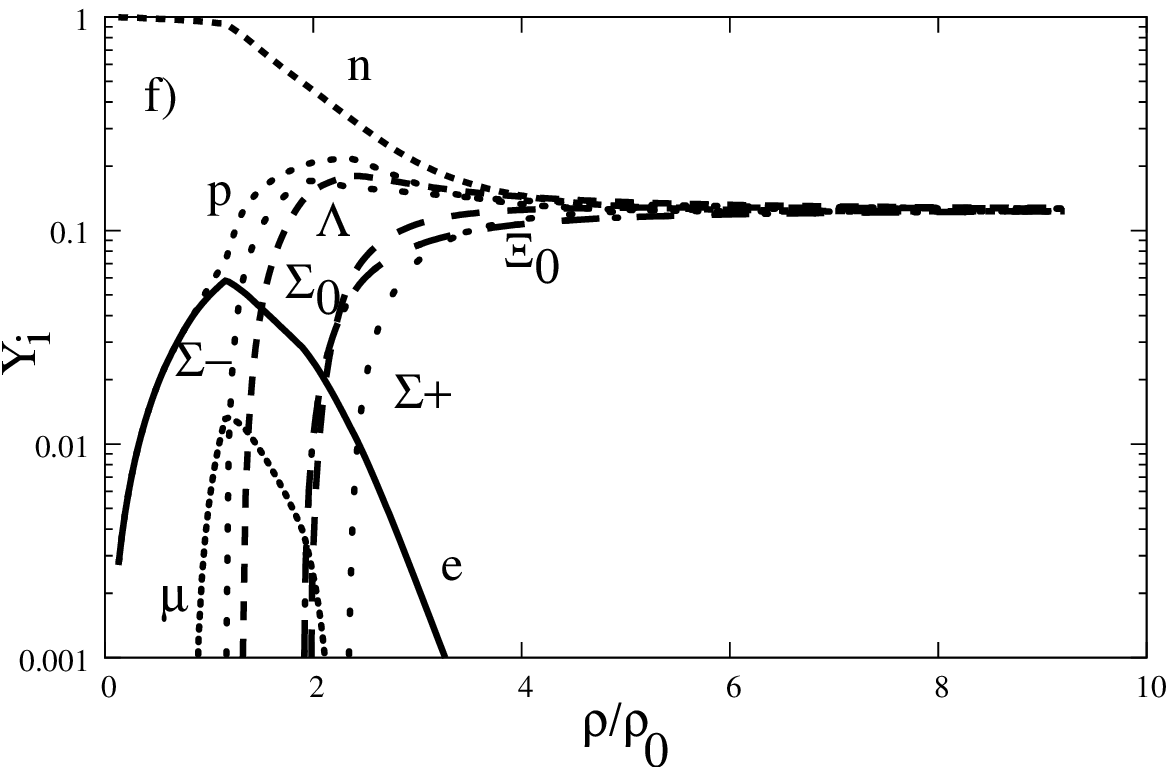} \\
\includegraphics[width=7.2cm,angle=0]{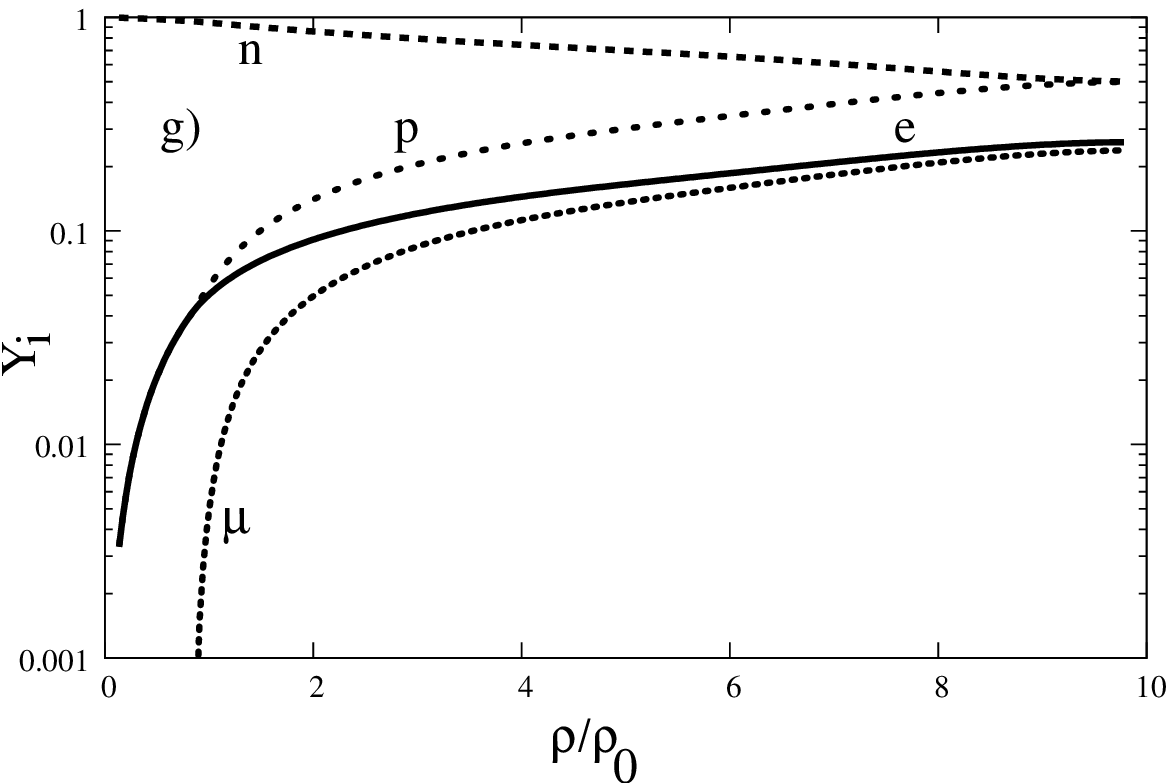} &
\includegraphics[width=7.2cm,angle=0]{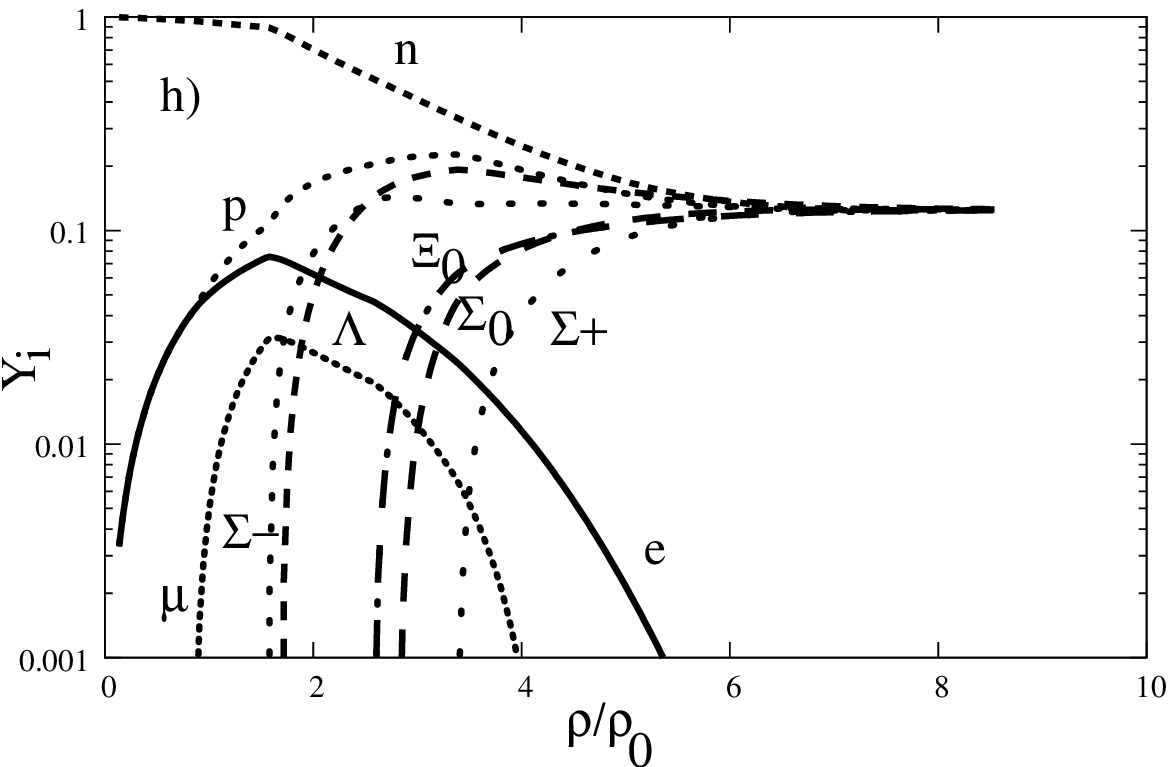} \\
\end{tabular}
\caption{Particle fractions for a) the non linear Walecka model with the GL
 parametrization with nucleons only, b) with 8 baryons,
c) TW model with nucleons only, d) with 8 baryons,
e) BR model with parameter sets S3 and nucleons only f) S3 and 8 baryons,
g)MA and nucleons only, h) MA and 8 baryons.}
\label{fraction}
\end{figure}

\begin{figure}
\begin{tabular}{cc}
\includegraphics[width=10.cm,angle=0]{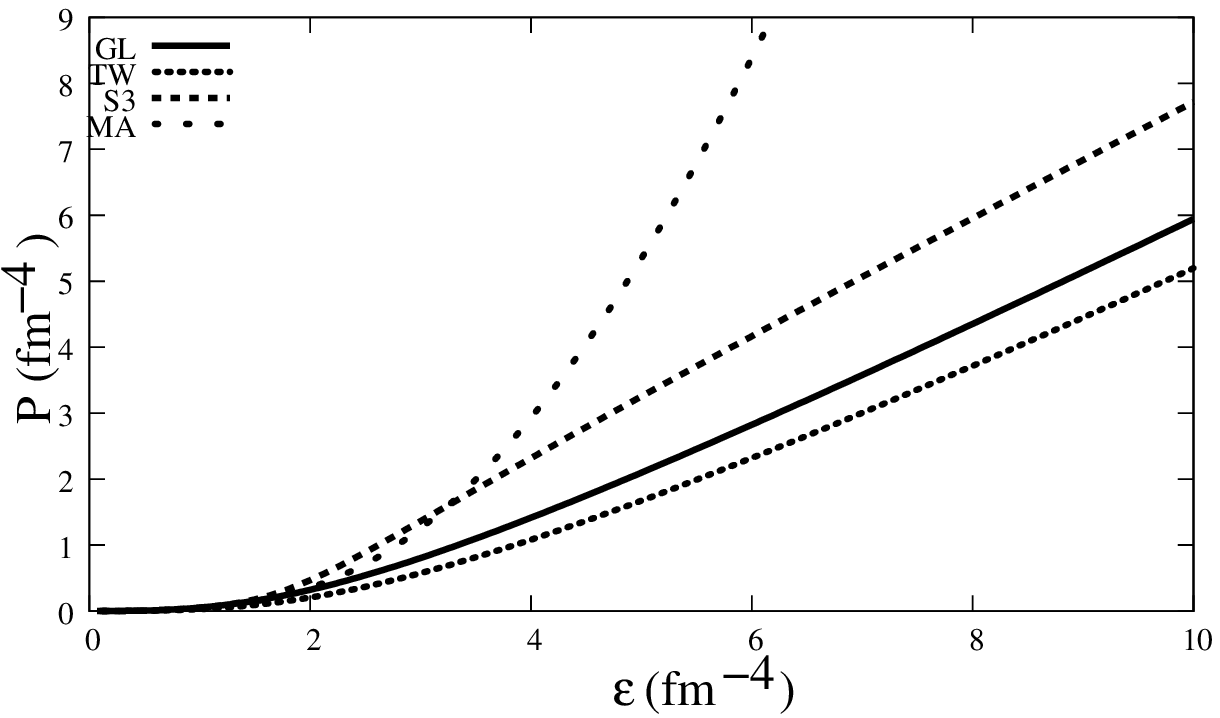} \\
\includegraphics[width=10.cm,angle=0]{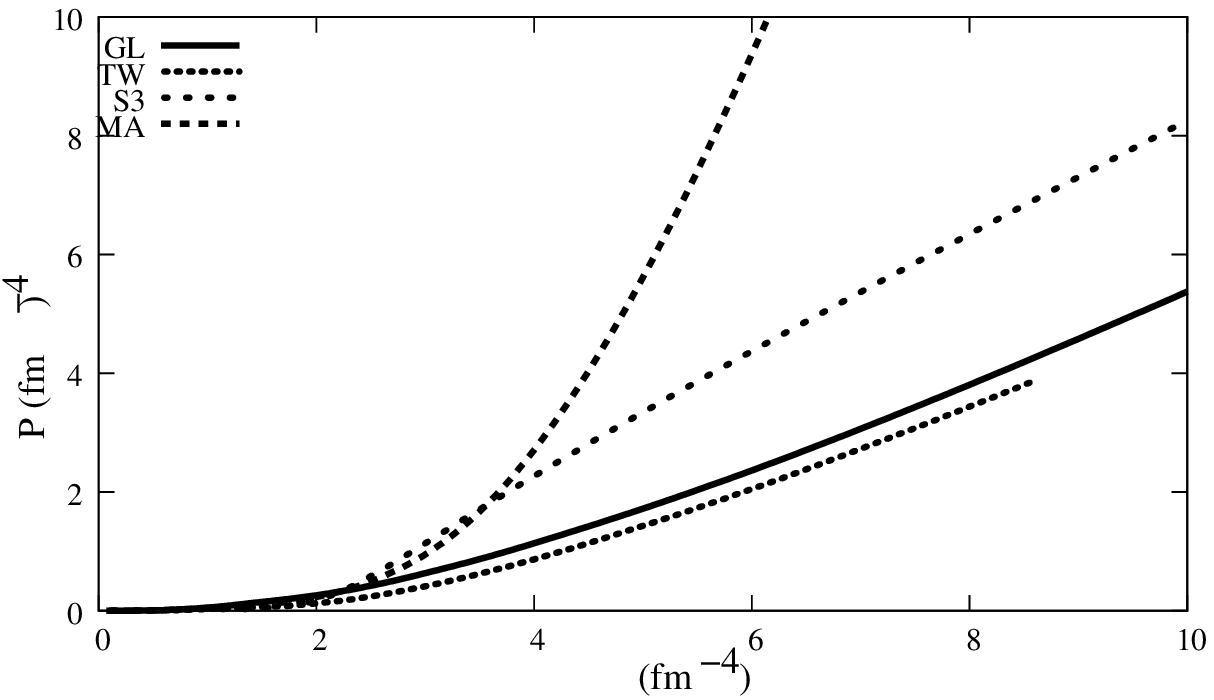}\\
\end{tabular}
\caption{EOS for the models discussed in the present work with nucleons
only (top figure) and 8 baryons (bottom figure).} 
\label{figeos}
\end{figure}

 In figure \ref{fraction} we show the particle population for the NLWM with
the GL parametrization, for the TW and the BR scaled model either with nucleons
only or with the 8 baryons. The particle fraction is defined as 
$Y_i=\rho_i/\rho$, $i=8$ baryons and 2 leptons.
We have again chosen two possibilities for the
BR scaled parameters, namely S3 and MA. If just protons and neutrons are 
included, the TW model presents a slight decrease in protons and consequently 
slight increase in neutrons and the MA model shows the opposite behavior as
compared with the GL parametrization. If hyperons are considered the results 
are all quite different, as a consequence of the different EOS shown in figure
\ref{figeos}. One can clearly see that all surviving particles tend to the
same amount in the BR scaled model, probably a consequence of the enforced
scaling law, a feature which happens earlier within S3 than with MA.
From figure \ref{figeos}, one can see that in both cases, the TW EOS 
is the softest one and the MA the hardest. S3 and GL interpolate between the 
other two EOS. For the present
choice of parameters, the TW model also stops 
converging at a too low density for astrophysical studies. The consequences of 
this fact will be discussed next.

 Once the EOS are obtained, we solve the Tolman-Oppenheimer-Volkoff
equations \cite{tov} in order to obtain the stellar properties.
They read
\begin{equation}
\frac{dP}{dr}=-\frac{G}{r}\frac{\left[{\cal E}+P\right ]\left[M+
4\pi r^3 P\right ]}{(r-2 GM)},
\label{tov1}
\end{equation}
\begin{equation}
\frac{dM}{dr}= 4\pi r^2 {\cal E} ,
\label{tov2}
\end{equation}
with $G$ as the gravitational constant and $M(r)$ as the enclosed gravitational
mass. We have used $c=1$.
Given an EOS, these equations can be integrated from the origin as an initial
value problem for a given choice of the central energy density, 
$(\varepsilon_0)$.
The value of $r~(=R)$, where the pressure vanishes defines the
surface of the star. 
 In Table \ref{tab3} we display the results for the stars with the 
maximum gravitational mass, the maximum baryonic mass, their radii and central
energy density for each of the EOS discussed in the present work.
One can see that, if only nucleons are considered, the TW model presents
the lowest maximum mass and the smallest radius with a consequent very large
central energy density. If hyperons are included the obtained result for the
TW model is just shown for completeness because it is unrealistic once the
maximum mass was not achieved since the program failed to converge at high
densities. The results obtained for the maximum masses with the BR scaled
models are somewhat larger than with the GL model, but still in the expected
range of values. Different results can be obtained with a different choice
of the $x_{s B}$, $x_{v B}$ and $x_{\rho B}$ constants.
Once hyperons are included, the EOSs always become softer with
a consequent lower value for the maximum stellar masses and radii and larger
central energy density.

 \begin{table}[ht]
\caption{Hadronic star properties for the EOSs
described in the text}
\begin{ruledtabular}
\begin{tabular}{lccccccc}
type & hadron model & $M_{max}(M_\odot)$ &
$M_{b~max}(M_\odot)$ & $R$ (Km) & $\varepsilon_0$ (fm$^{-4}$)\\
\hline
np               & GL  & 2.40 & 2.89 & 12.19 & 5.43 \\
np+hyperons      & GL  & 2.18 & 2.56 & 11.35 & 6.34 \\
np               & TW  & 2.08 & 2.46 & 10.62 & 7.20 \\
np+hyperons(*)   & TW  & 1.89 & 2.24 &  9.46 & 8.44 \\          
np               & S3  & 2.88 & 3.57 & 12.81 & 4.57 \\
np+hyperons      & S3  & 2.65 & 3.36 & 11.33 & 5.60 \\
np               & MA  & 2.86 & 3.59 & 11.79 & 5.29 \\
np+hyperons      & MA  & 2.76 & 3.49 & 11.00 & 5.93 \\
\hline
\end{tabular}
\end{ruledtabular}
\label{tab3}
\end{table}

  \section{The importance of the rearrangement terms}

 In what follows we concentrate on the role played by the rearrangement term
in the different models. This term is very important and influences all
properties of nuclear and neutron star matter since it changes the effective
chemical potentials given in equations (\ref{efchem}),(\ref{effbr}) and
(\ref{efftw}).

\begin{figure}
\begin{tabular}{cc}
\includegraphics[width=7.5cm,angle=0]{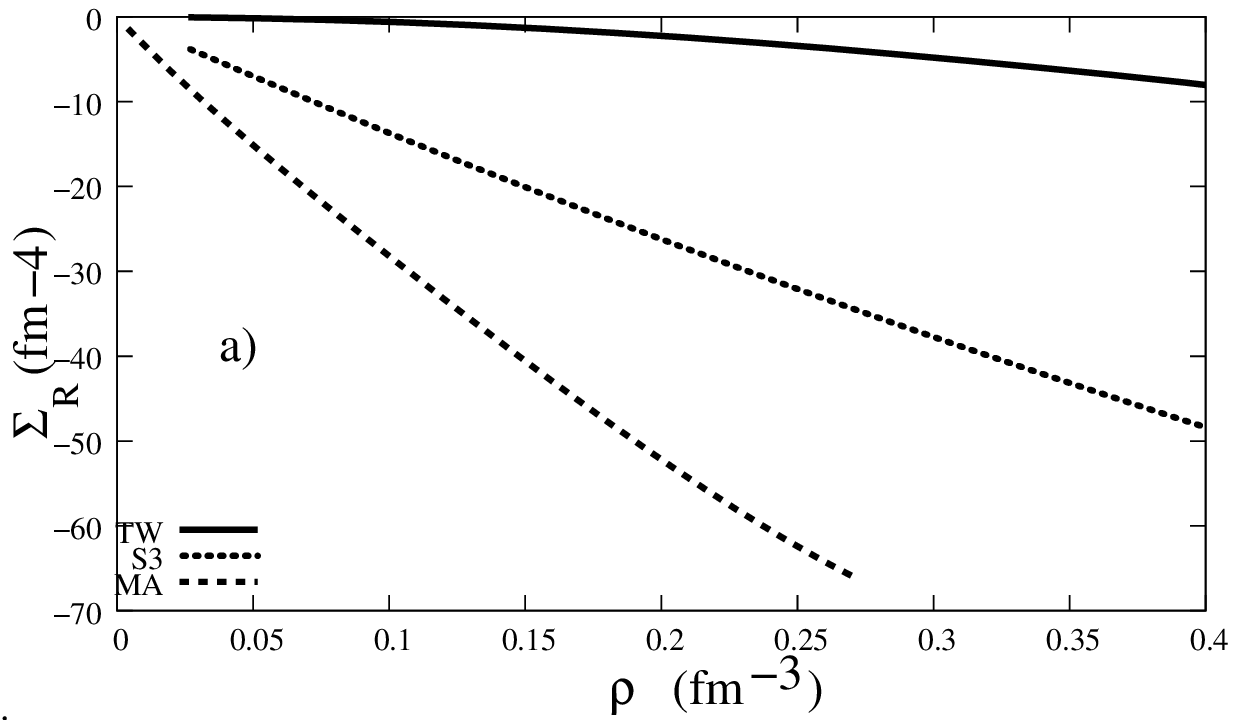} &
\includegraphics[width=7.5cm,angle=0]{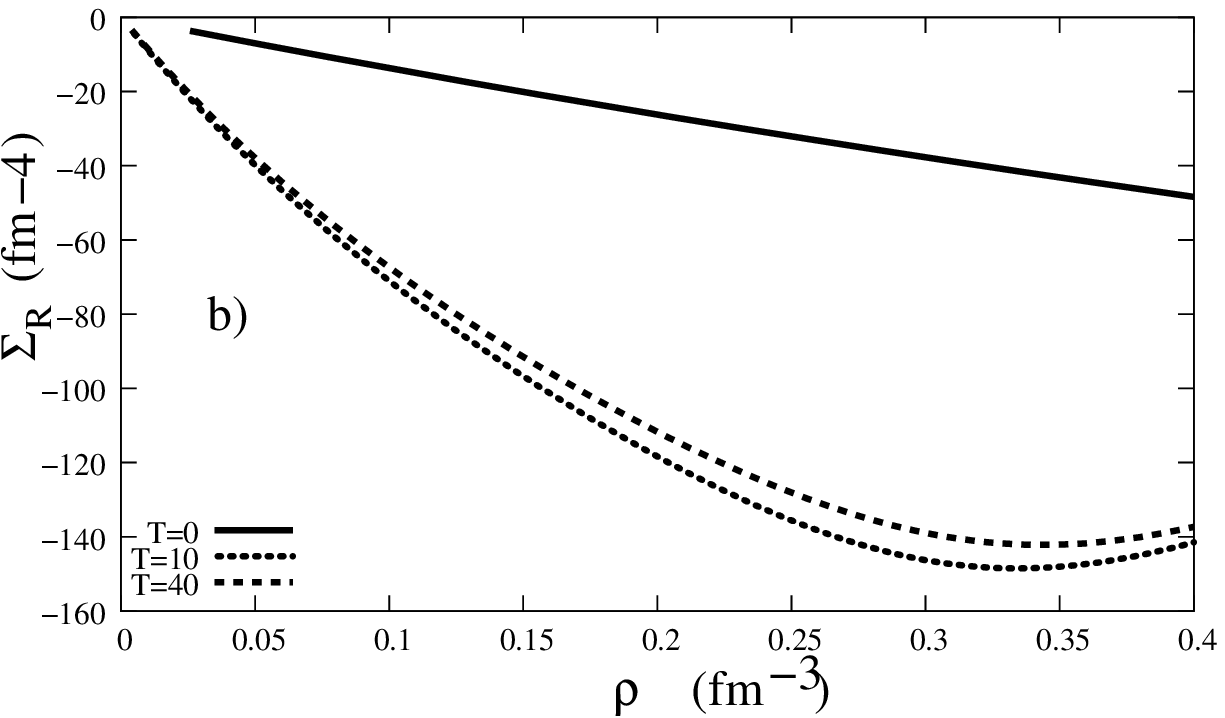}\\
\includegraphics[width=7.5cm,angle=0]{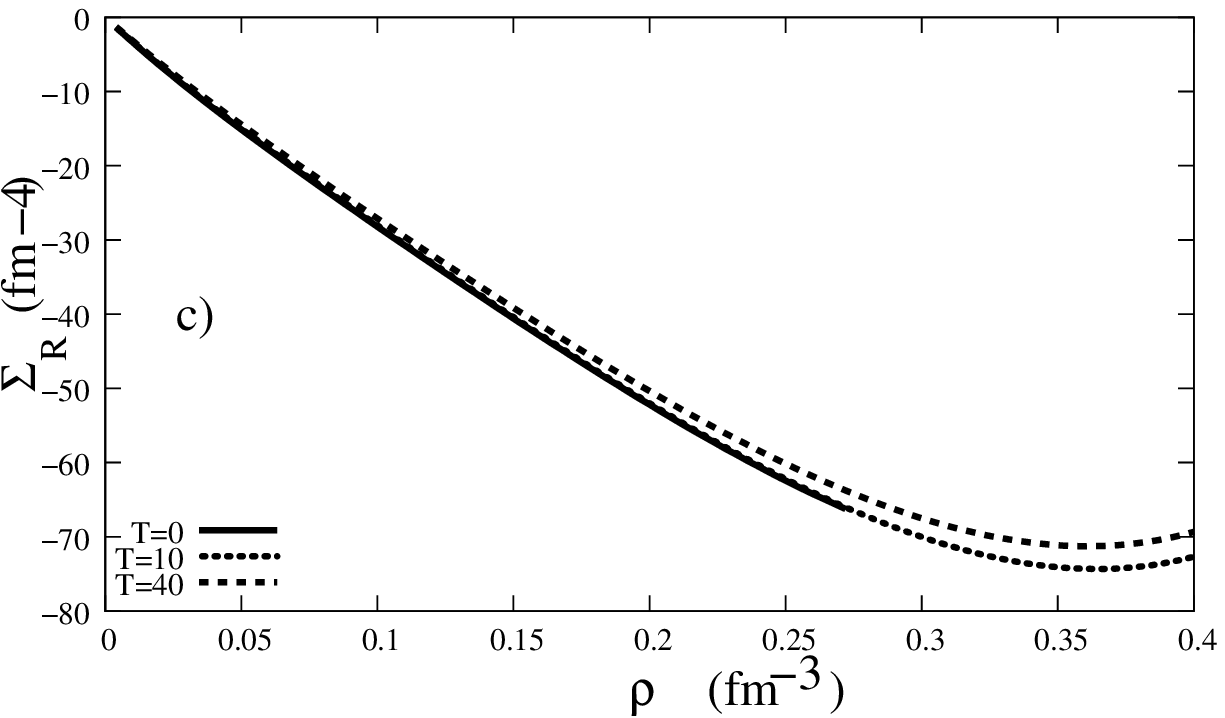}&
\includegraphics[width=7.5cm,angle=0]{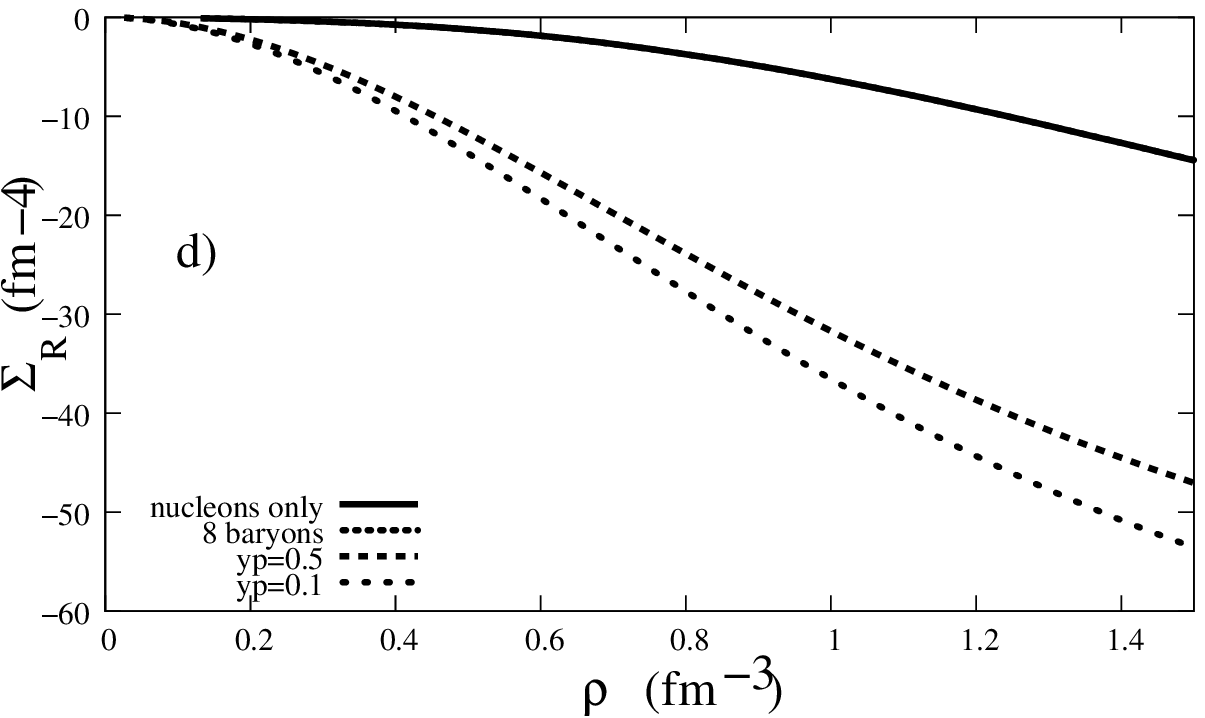}\\
\includegraphics[width=7.5cm,angle=0]{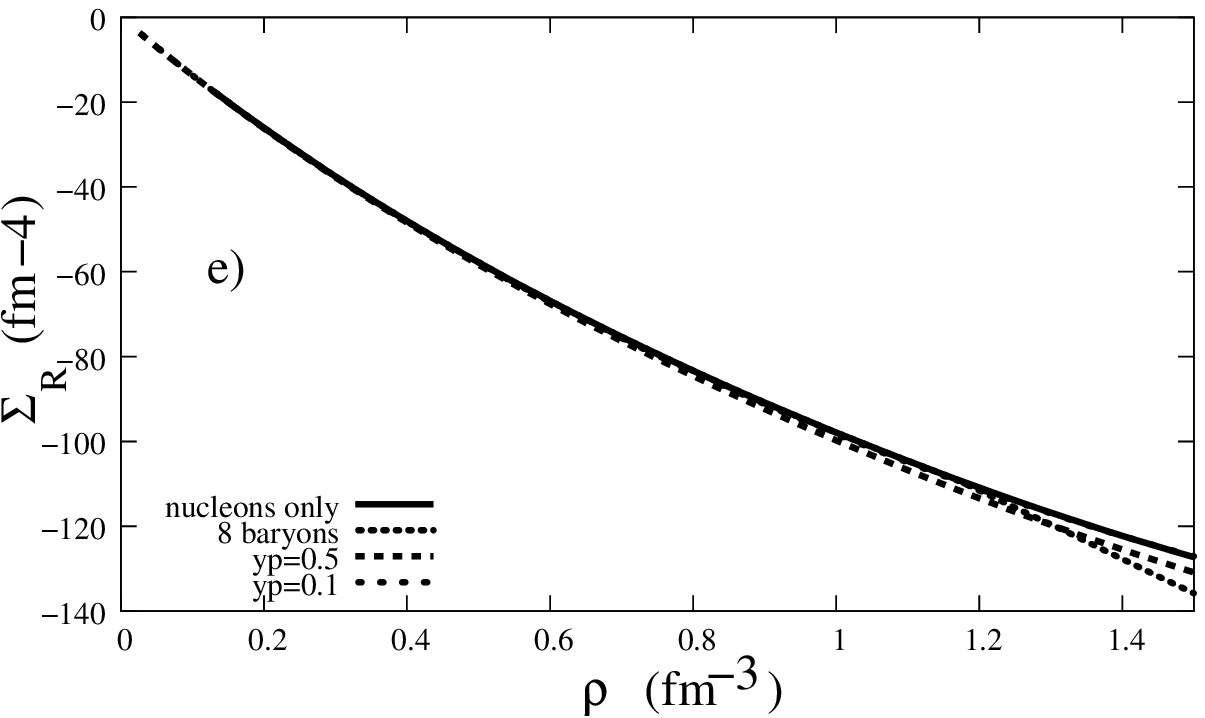}&
\includegraphics[width=7.5cm,angle=0]{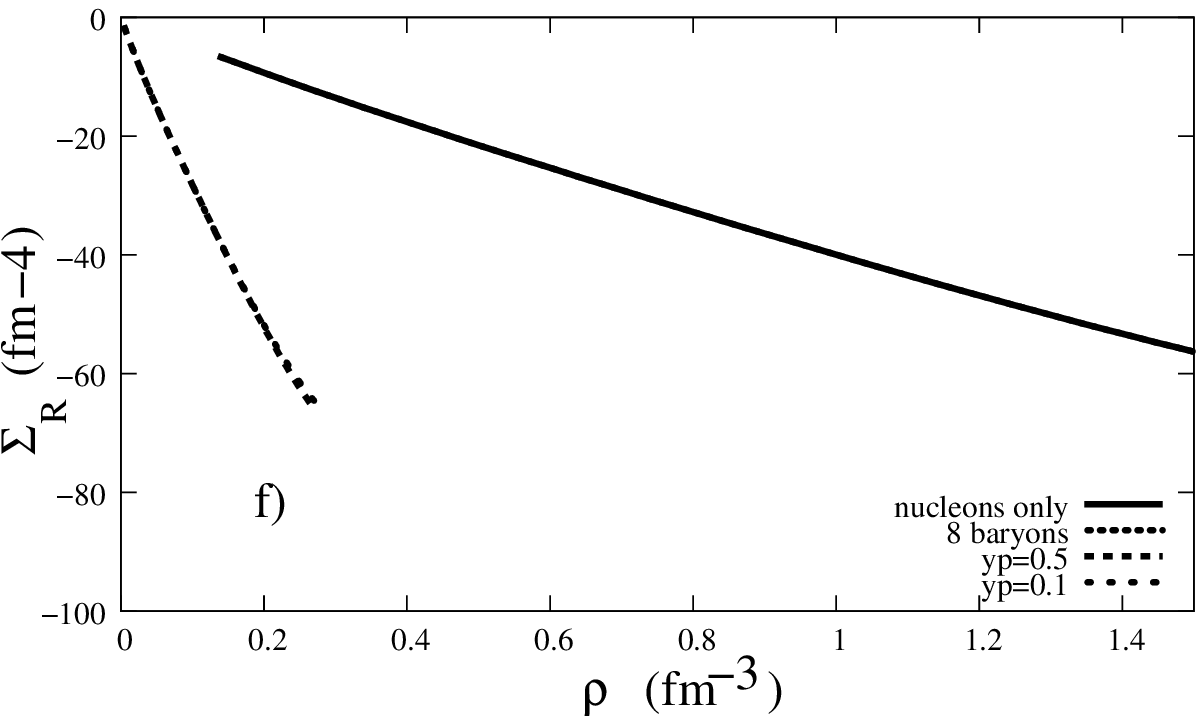}\\
\includegraphics[width=7.5cm,angle=0]{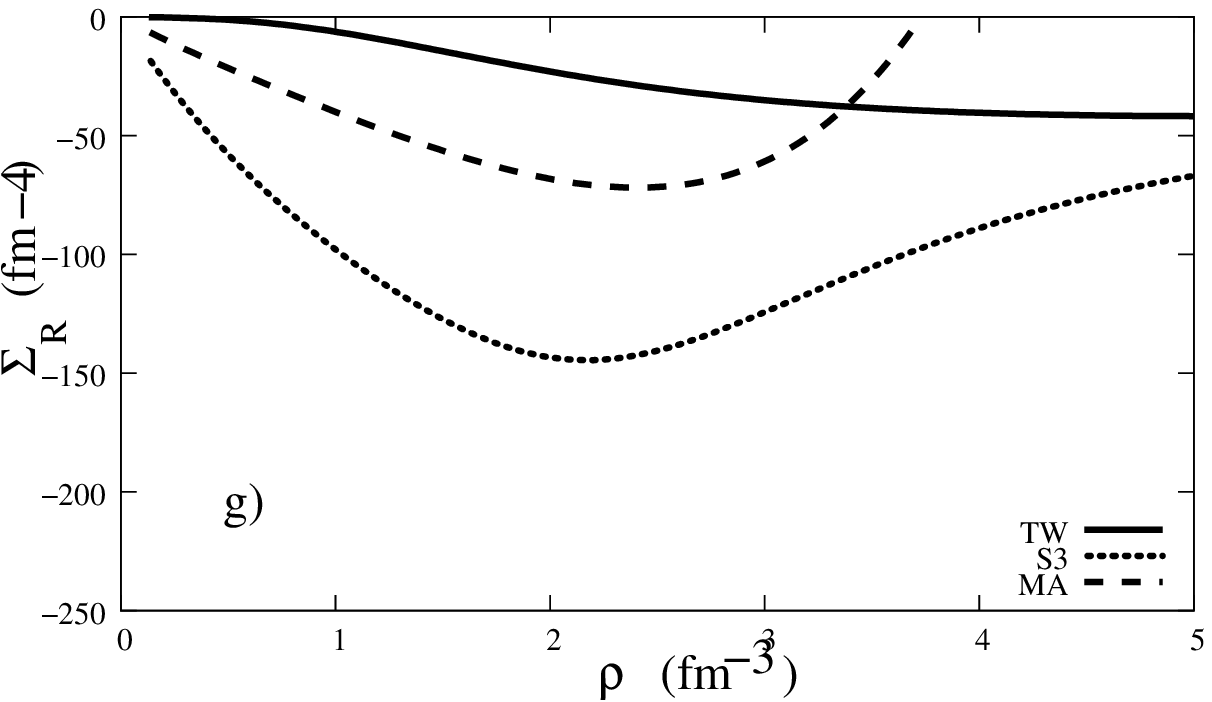} & 
\includegraphics[width=7.5cm,angle=0]{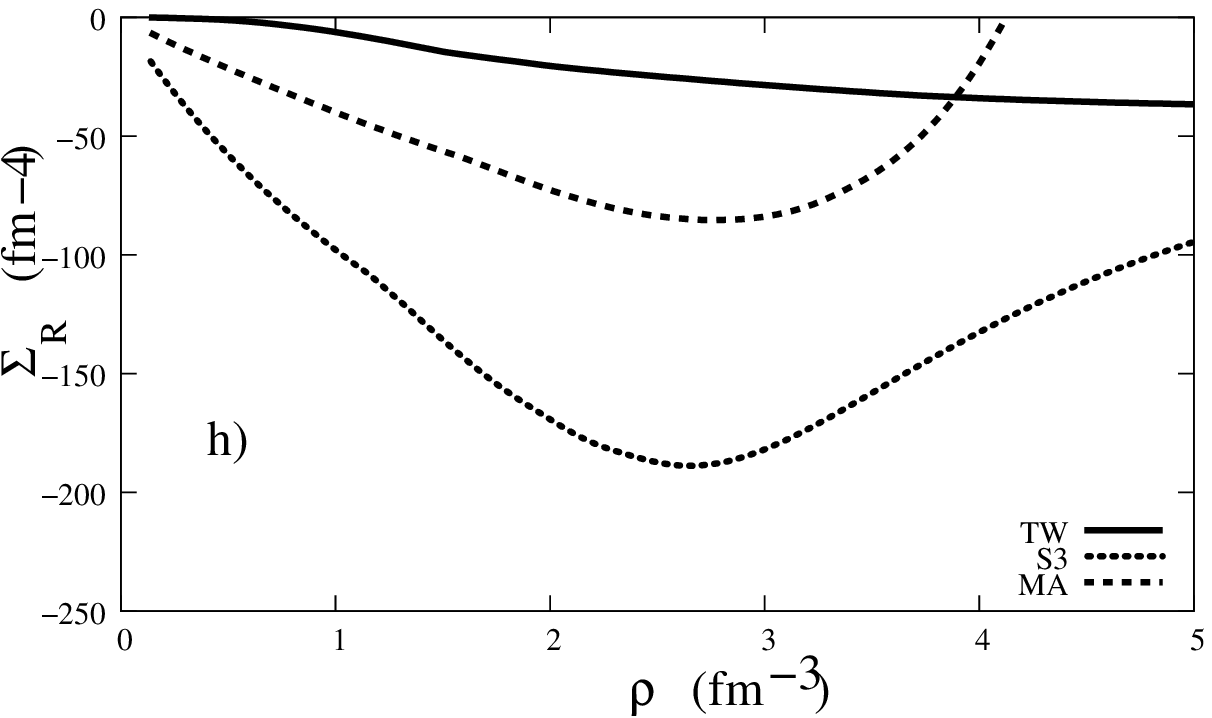}\\
\end{tabular}
\caption{Rearrangement terms calculated in different situations:
a)different models in symmetric nuclear matter, b) S3 and symmetric nuclear 
matter for different temperatures, c) MA and symmetric nuclear matter for 
different temperatures, d) TW, e) S3, f) MA, g) nucleons only, h) 8baryons.}
\label{figrearr}
\end{figure}

 In figure \ref{figrearr} we show the rearrangement term which arises in
various situations. In figure \ref{figrearr}a, it is shown the rearrangement
terms which appear in symmetric nuclear matter (without the imposition of
$\beta$ stability) for the TW and the two parametrizations of the BR scaled
model within the density range considered in nuclear matter studies.
The S3 parametrization produces a term which is much more negative
than the TW model. The MA parametrization shows a rearrangement term which 
decreases even further. So, in this case the rearrangement term is more 
attractive within the BR-scaled models.
Notice that in the S3 parametrization, if $y$ and 
$z$ were chosen as having the same value, only the scalar meson would 
contribute to the rearrangement term.

Please, notice that the scales of the figures mentioned next are all different.
In figures \ref{figrearr}b and \ref{figrearr}c we show the influence of the
temperature on the rearrangement term of the two BR scaled models. As the
temperature affects very little the term in the MA parametrization, the same
does not happen if the S3 parametrization is chosen. Moreover, the 
rearrangement term increases slightly as temperature increases within the MA 
and decreases quite a lot with the increase of temperature with the S3 choice.

 In figure \ref{figrearr}d, the
rearrangement term of the TW model is shown for symmetric nuclear matter, for
a very asymmetric nuclear matter, with a proton fraction 
$y_p=0.1$ and for the equation of state where charge neutrality
and $\beta$ stability is required either with only protons and neutrons or with
hyperons as well. The same is shown in figures \ref{figrearr}e and 
\ref{figrearr}f for S3 and MA. While in the TW model the rearrangement term
decreases with the asymmetry of the system and increases when the conditions
of $\beta$ equilibrium and charge neutrality are enforced, the influence of 
the hyperons being very small, in the S3 all curves are very similar and in 
the MA the asymmetry almost does not interfere in the rearrangement term
and the hyperons again play no role, but charge neutrality and $\beta$ 
equilibrium conditions modify the curve quite drastically.

 In figures \ref{figrearr}g and \ref{figrearr}h one
can see a comparison between the rearrangement term arising from the TW model
and the ones obtained with the BR scaled models respectively for the case
when only nucleons are considered and when hyperons are also included in the 
system. We have now opted to show a much wider density range so that the
differences can be clearly seen. While the inclusion of hyperons makes the 
TW model rearrangement term increase slightly at densities of the order of 
$\simeq 2$ fm$^{-3}$ as compared with the case when only nucleons are 
considered, at about the same density the MA rearrangement term
starts to decrease. The S3 term also decreases more rapidly if hyperons are 
considered, but it starts at a density of
the order of $\simeq 1$ fm$^{-3}$. At these very large densities other 
important features as the deconfinement to the quark matter is already present.

 \section{Conclusions}

 In the present work we have derived a formalism to incorporate temperature 
effects in the BR scaled model to make it useful in future heavy ion 
collision studies. We have also investigated the possibility of applying
it to nuclear astrophysics by enforcing charge neutrality and $\beta$ 
equilibrium. We have compared our results with the more standard NL3 \cite{nl3}
version of the NLWM in nuclear matter and with the GL \cite{Glen00} 
parametrization of the NLWM in neutron star matter. In both cases, the BR 
scaled model was also
compared with another density dependent model, the TW \cite{tw}. It is worth 
 pointing out that density dependent models are alternative approaches to 
describe hadronic matter without the usual non-linear terms necessary in the
NLWM. Although in the low baryonic density regime all models and 
parametrizations used in this paper have EOSs with similar behavior, 
different scenarios show up when the density increases, specially, when the 
baryon octet is taken into account.
The lagrangians with BR scaling are much simpler than the other model with 
density dependent couplings (TW) and they provide effective meson
masses which decrease with the increase of the density, behavior which has
already been confirmed by experiments \cite{taps,naruki} 
and which is not present in all models where the meson masses are held fixed.
 We have also shown that the BR scaled model is a very good tool in describing
neutron star properties. Of course a more systematic study can be done by
including the delta \cite{deltas,alemaes} and the strange 
\cite{rafael,alemaes} mesons in the lagrangian density. Other possible choices 
for the meson-hyperon couplings should also be considered. More
realistic hybrid stars with a deconfinement to the quark phase can now be 
obtained within the BR scaled model for the hadron phase. When the new EOSs 
are built and the stellar properties are obtained, the mass to radius 
constraints \cite{cottam,sanwal} can be used as a probe to the new formalism.
 It is also important to say that the rearrangement term plays a central
role in density dependent models and many of the system properties depend on 
its strength.It is crucial to assure the energy-momentum conservation and 
the thermodynamical consistency for density dependent models. In contrast to 
the pressure and compressibility which depends explicitly on the rearrangement 
term, in the energy density it cancels out. However, it still contributes to 
the binding energy through the chemical potentials. In figure \ref{figrearr} 
it was shown that the BR parametrizations yields more attractive
rearrangement term as compared with the TW model.
 
  We have checked that other possibilities for the parametrization of the BR 
scaled model given by
\beq
\frac{g_s^*}{g_s}=\frac{1+x_1 \rho/\rho_0}{1+x_2 \rho/\rho_0},~~
\frac{g_v^*}{g_v}=\frac{1+x_3 \rho/\rho_0}{1+x_4 \rho/\rho_0},~~
\frac{g_\rho^*}{g_\rho}=\frac{1+x_5 \rho/\rho_0}{1+x_6 \rho/\rho_0},
\label{maisuma}
\eeq
also work. Nevertheless, one of the advantages of using the BR scaled models
instead of the TW model is that they contain a much smaller number of 
parameters. With the choice given in (\ref{maisuma}) much of the beauty of
the model would be lost, but it remains as a possible alternative.
Finally, it is important to stress that a test to finite nuclei in order
to obtain the correct value for the spin-orbit splitting strength is still
necessary with the BR parameter sets used in this work. 
 
 \section*{ACKNOWLEDGEMENTS}

This work was partially supported by CNPq (Brazil) and Capes (Brazil) under 
process BEX 1681/04-4.
D.P.M. would like to thank the friendly atmosphere at the Reserch Centre
for Theoretical Astrophysics, Sydney University, where this
work was partially done.

\end{document}